\documentclass[%
reprint,
superscriptaddress,
 amsmath,amssymb,
 aps,
pre,
floatfix,
]{revtex4-2}

\usepackage{graphicx}%
\usepackage{bm}%
\usepackage{hyperref}
\usepackage{units}
\usepackage{ifthen}
\usepackage{xcolor}

\DeclareMathOperator{\STD}{SD}
\DeclareMathOperator{\SEM}{SEM}

\newcommand{\mean}[1]{\langle #1 \rangle}

\newcommand{\kBT}{k_\mathrm{B} T}
\newcommand{\Eqref}[1]{Eq.~\eqref{#1}}
\newcommand{\Eqsref}[1]{Eqs.~\eqref{#1}}
\newcommand{\figref}[1]{Fig.~\ref{#1}}
\newcommand{\diff}{\mathrm{d}}

\newcommand{\p}[1][]{\ifthenelse{\equal{#1}{}}{{\phi}}{\phi^{(#1)}}}
\newcommand{\m}[1][]{\ifthenelse{\equal{#1}{}}{{\mu}}{\mu^{(#1)}}}
\newcommand{\nm}[1][]{\ifthenelse{\equal{#1}{}}{{\hat\mu}}{\hat\mu^{(#1)}}}
\newcommand{\nP}[1][]{\ifthenelse{\equal{#1}{}}{{\hat P}}{\hat P^{(#1)}}}
\newcommand{\V}[1][]{\ifthenelse{\equal{#1}{}}{{V}}{V^{(#1)}}}
\newcommand{\N}[1][]{\ifthenelse{\equal{#1}{}}{{N}}{N^{(#1)}}}
\newcommand{\phiS}[1][]{\ifthenelse{\equal{#1}{}}{{\phi_0}}{\phi_0^{(#1)}}}
\newcommand{\sigmaE}{\sigma_\mathrm{e}}
\newcommand{\sigmaP}{\sigma_\mathrm{p}}
\newcommand{\chiBound}{\chi_\mathrm{bound}}

\newcommand{\x}[1][]{\ifthenelse{\equal{#1}{}}{{x}}{x^{(#1)}}}
\newcommand{\y}[1][]{\ifthenelse{\equal{#1}{}}{{y}}{y^{(#1)}}}
\newcommand{\partialT}{\partial_{\hat t}}

\newcommand{\bigzero}{\mbox{\normalfont\Large\bfseries 0}}

\setcitestyle{numbers,square,sort&compress}

\begin{document}

\title{Evolved interactions stabilize many coexisting phases in multicomponent liquids}

\author{David Zwicker}
\email{david.zwicker@ds.mpg.de}
\affiliation{Max Planck Institute for Dynamics and Self-Organisation, Göttingen, Germany}%

\author{Liedewij Laan}
%\email{david.zwicker@ds.mpg.de}
\affiliation{Department of Bionanoscience, TU Delft, 2629 HZ Delft, The Netherlands}%

%\correspondingauthor{\textsuperscript{1}To whom correspondence should be addressed. E-mail: david.zwicker@ds.mpg.de}

\begin{abstract}
Phase separation has emerged as an essential concept for the spatial organization inside biological cells.
However, despite the clear relevance to virtually all physiological functions, we understand surprisingly little about what phases form in a system of many interacting components, like in cells.
Here, we introduce a new numerical method based on physical relaxation dynamics to study the coexisting phases in such systems.
We use our approach to optimize interactions between components, similar to how evolution might have optimized the interactions of proteins.
These evolved interactions robustly lead to a defined number of phases, despite substantial uncertainties in the initial composition, while random or designed interactions perform much worse.
Moreover, the optimized interactions are robust to perturbations and they allow fast adaption to new target phase counts.
We thus show that genetically encoded interactions of proteins provide versatile control of phase behavior.
The phases forming in our system are also a concrete example of a robust emergent property that does not rely on fine-tuning the parameters of individual constituents.
\end{abstract}

\maketitle
%\tableofcontents

\newpage

Biological cells are incredibly complex and consist of thousands of different biomolecules that move and react rapidly.
Yet, cells display robust behavior, partly because they separate molecules into distinct compartments.
One important class of compartments are biomolecular condensates, which have now been identified in  eukaryotes~\cite{Brangwynne2009, Feric2016, Banani2017}, procaryotes~\cite{Azaldegui2020, Cohan2020, Greening2020}, and plants~\cite{Emenecker2021, Kim2021}.
In all systems, multiple different condensates coexist and some condensates, like the nucleolus~\cite{Lafontaine2020} and nuclear speckles~\cite{Fei2017}, even possess sub-compartments.
The collective organization of biomolecules into condensates is explained by phase separation~\cite{Fritsch2021}, which is a physical mechanism where a gain in enthalpic interactions offsets the entropy loss when molecules are confined.
Since all proteins interact weakly by various mechanisms~\cite{Dignon2020}, phase separation is widely expected in the proteome~\cite{Hardenberg2020} and transcriptome~\cite{Adekunle2020}.
However, it is still mysterious how cells regulate phase separation.

Biomolecular condensates need to form robustly, despite internal and external uncertainties that cells cannot control.
Having the right condensates, in the right situation, at the right time is crucial since condensates participate in almost all cellular processes~\cite{Lyon2020}, they affect the fitness of prokaryotes~\cite{Jin2021}, and malfunctioning is implicated in many diseases~\cite{Alberti2019a}.
It is particularly mysterious how cells reliably form many different kinds of condensates in a common cytosol, despite copy number fluctuations of all components. %
Are the interactions between components tuned such that the right condensates form reliably?
It is conceivable that multiple driving forces of phase separation~\cite{Dignon2020} have been adjusted over evolutionary time scales.
Indeed, theoretical studies~\cite{Saare2021,Choi2020a}, numerical simulations~\cite{Harmon2017, Lin2018}, and \textit{in vitro} experiments~\cite{Schuster2020,Bremer2021} demonstrated that small modifications of the sequence of a protein can have profound impact on its phase separation.
However, it is not clear whether these results on single components can be transferred to multicomponent mixtures.

While the theoretical basis of phase separation is well-understood~\cite{Hyman2014,Brangwynne2015,Berry2018, Weber2019}, even predicting equilibrium states is challenging in multicomponent mixtures.
This is due to enormous variability in heterotypic interactions, which leads to complex phase diagrams~\cite{Riback2020}.
We can now construct complete phase diagrams for up to $5$ components~\cite{Mao2018} and predict the associated phase morphology~\cite{Mao2020}. %
This showed that the number of coexisting phases typically depends on the overall composition of the system, but it is unclear how this phase count depends on the component count and the specific interaction matrix.
Answering this question is critical, since typical biological condensates consist of many components~\cite{Leung2003,Updike2009, Riback2020, Currie2021} and the scaffold-client picture~\cite{Banani2016}, where a single scaffold component dominates the phase behavior, might not always apply.
State-of-the art numerical techniques can simulate mixtures of up to $16$ components~\cite{Zhou2021a, Shrinivas2021}, but these techniques are often too costly to truly explore the space of possible interactions.
Random matrix theory provides an alternative approach to investigate the stability of mixtures comprising very many components whose interactions are chosen from a random distribution~\cite{Sear2003,Jacobs2013,Jacobs2017,Shrinivas2021,Jacobs2021}.
While these studies demonstrated that phase separation is overwhelmingly likely in such systems, it is unclear how well random interactions capture real proteins, which have evolved for millions of generations.
In fact, it is unclear what properties of interacting proteins need to be conserved during evolution for a robust phase separation behavior.

\section*{Results}

We here present a novel approach to analyze multiphase equilibrium states of multicomponent liquids, which is based on relaxation dynamics.
We then use this model to investigate how  components need to interact such that a given number of phases forms reliably.

\subsection*{A simplified physical model reveals equilibrium states}

\begin{figure}
	\centering	
	\includegraphics[width=\columnwidth]{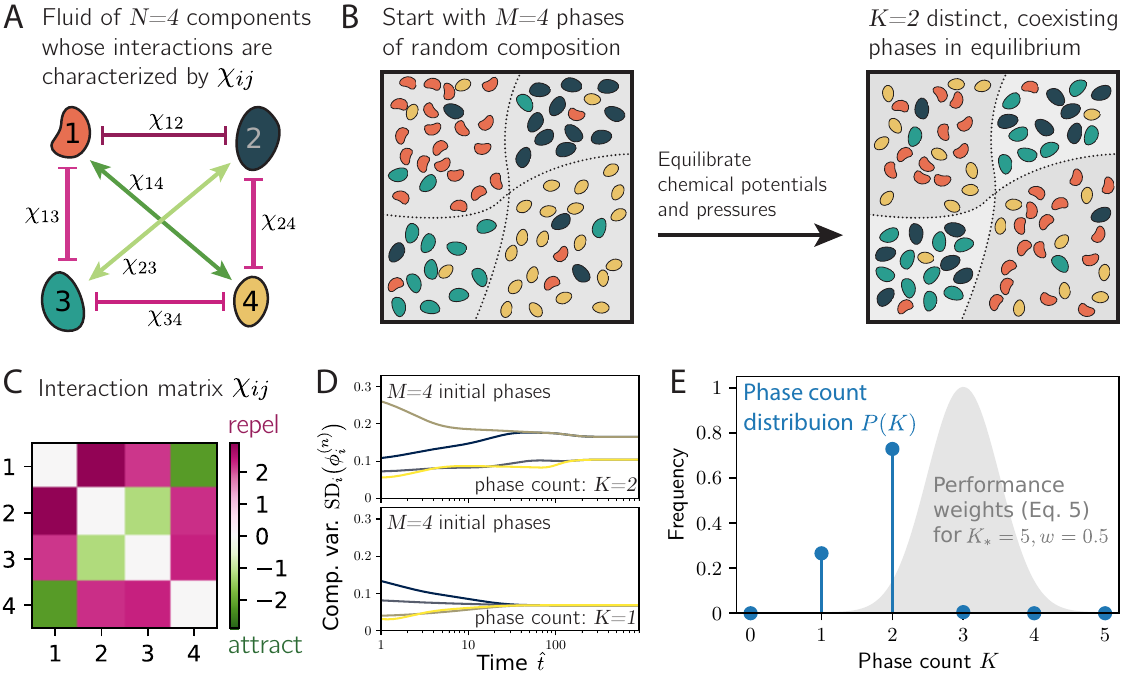}
	\caption{
	A dynamical system recovers coexisting phases of the multicomponent liquid.
	(A)
	Schematic of $N=4$ components with attractive ($\chi_{ij} < 0$, green arrows between orange/red and teal/gray components) and repulsive ($\chi_{ij} > 0$, pink lines between remaining pairs) interactions.
	(B)
	Schematic showing how the $N$-component liquid is initially split into $M=4$ phases of random composition.
	After equilibrating chemical potentials and pressures, only $K=2$ phases of distinct composition remain.
	(C) Interaction matrix~$\chi_{ij}$ corresponding to panel A.
	(D) Two representative simulations with different initial composition for $\chi_{ij}$ of panel C.
	The composition variation $[\mean{(\p[n]_i)^2}_i - \mean{\p[n]_i}_i^2]^{1/2}$ is shown as a function of time~$\hat t$.
	(E) Frequencies $P(K)$ of phase counts $K$ for random initial conditions.
	The performance~$g$ follows from a convolution of $P(K)$ with weights (gray area); see \Eqref{eqn:performance}.
	}
	\label{fig:simulations}
\end{figure}

We consider an isothermal, incompressible liquid comprised of $N$ different components and an inert solvent.
In equilibrium, such a system can in principle form $N+1$ liquid phases~\cite{Gibbs1876}, which are homogeneous regions with distinct composition.
However, in typical realistic systems fewer phases form since some components might be miscible.
To reveal how the number of phases formed depends on the interactions of the components, we consider the general case of $M$ coexisting phases with volumes~$\V[n]$ for $n=1,\ldots,M$.
Since phases are homogeneous, their composition is fully described by the particle counts~$\N[n]_i$ for each component $i=1,\ldots,N$ or the associated volume fractions~$\p[n]_i = \nu \N[n]_i/\V[n]$, where we consider equal molecular volumes $\nu$ for simplicity.
Note that the fraction of the inert solvent, $\phiS[n] = 1 - \sum_{i=1}^N \p[n]_i$, is not an independent variable.
Multiple phases can coexist when the associated free energy $F=\sum_{n=1}^M \V[n]f(\{\p[n]_i\})$ is minimal, where $f$ is the free energy density that depends on the local composition.
We here consider regular solution theory,~\cite{Flory1942}
\begin{equation}
	\label{eqn:free_energy}
	f(\{\phi_i\}) = \frac{\kBT}{\nu}\biggl[
		\phiS\ln(\phiS)
		+ \sum_{i=1}^N \phi_i \ln(\phi_i)
		+ \!\sum_{i,j=1}^N \frac{\chi_{ij}}{2} \phi_i\phi_j
	\biggr]
	\;,
\end{equation}
where $\kBT$ is the thermal energy scale and the first two terms capture the entropic contributions of the solvent and all other components, respectively.
Conversely, the last term quantifies the enthalpic interaction between all components.
The elements of the interaction matrix~$\chi_{ij}$ can for instance be derived from the interaction energies~$w_{ij}$ between components $i$ and $j$ on a lattice, $\chi_{ij} =  z(2w_{ij} - w_{ii} - w_{jj})/(2\kBT)$, where $z$ is the lattice coordination number~\cite{Cahn1958, Mao2018}.
This implies that the diagonal entries vanish, $\chi_{ii}=0$, while the off-diagonal entries capture the relevant balance between heterotypic and homotypic interactions; see \figref{fig:simulations}A.
Note that effective repulsion ($\chi_{ij}>0$) can originate not only from heterotypic repulsion ($w_{ij}>0$), but also from homotypic attraction that outweighs the heterotypic interaction ($w_{ii} + w_{jj} < 2w_{ij}$).

The multicomponent liquid reaches equilibrium when $F$ is minimal, implying that the chemical potentials~$\mu_i=  \nu \partial f/\partial \phi_i$ and the pressures $P =\sum_i \phi_i \partial f/\partial \phi_i - f$ are equal between all phases~\cite{Weber2019}.
We express these quantities in non-dimensional form, $\nm_i=\mu_i/\kBT$ and $\nP=P\nu/\kBT$, for each phase~$n$,
\begin{subequations}
\label{eqn:mu_p}
\begin{align}
	\nm[n]_i &= 
		\ln(\p[n]_i) - \ln(\phiS[n]) + \sum_{j=1}^N \chi_{ij} \p[n]_j
\\
	\nP[n] &=
		- \ln(\phiS[n]) + \sum_{i,j=1}^N \frac{\chi_{ij}}{2} \p[n]_i\p[n]_j
	\;.
\end{align}
\end{subequations}
The equilibrium conditions for the system then read
\begin{subequations}
\label{eqn:equilibrium}
\begin{align}
	\nm[1]_i &= \nm[2]_i = \cdots = \nm[M]_i \qquad\text{and}
\\
	\nP[1] &= \nP[2] = \cdots = \nP[M]
	\;,
\end{align}
\end{subequations}
for $i=1,\ldots, N$, which are $(M-1)N$ and $M-1$ non-linear equations, respectively.
Additionally, there are $N$ equations for the conservation of particles, $\sum_n \N[n]_i = \text{const}$, and an equation for volume conservation, $\sum_n \V[n] = \text{const}$.
Taken together, these equations can in principle be solved for the $M$ volumes~$\V[n]$ and $NM$ particle counts $\N[n]_i$, although this is generally challenging~\cite{Mao2018}.

The equilibrium conditions~(\ref{eqn:equilibrium}) describe the local coexistence of phases of potentially different composition~$\p[n]_i$.
Since these conditions only involve the intensive quantities~$\p[n]_i$, coexisting volume fractions can be determined without specifying the extensive volumes~$\V[n]$, similar to the Maxwell construction in a binary system~\cite{Weber2019}.
In multicomponent systems, such equilibrium points correspond to the stationary state of a simple dynamical system,
\begin{equation}
	\label{eqn:odes}
	\partial_{\hat t} \p[n]_i = 
		\p[n]_i \sum_{m=1}^M  \left[
			\p[m]_i \bigl(\nm[m]_i - \nm[n]_i\bigr)
			+ \nP[m] - \nP[n]
		\right]
	\;,
\end{equation}
where $\hat t$ is a non-dimensional time and the interaction matrix~$\chi_{ij}$ is the only parameter; see \Eqref{eqn:mu_p}.
Clearly, \Eqref{eqn:odes} is at a stationary state, $\partial_{\hat t} \p[n]_i = 0$, when the equilibrium conditions~(\ref{eqn:equilibrium}) are obeyed.
We show in Appendix~\ref{sec:appendix_dynamical_system} that the converse is also true, %
so the relaxation dynamics given by \Eqref{eqn:odes} lead us to equilibrium states whose composition we can then analyze further.
All possible equilibrium states together form the binodal manifolds of the $N$-dimensional phase diagram.
Since these manifolds can be very complicated~\cite{Mao2018}, we for simplicity focus on the distribution of the number of distinct, coexisting phases, $K$.
\figref{fig:simulations}D shows two trajectories, revealing the typical situation that some phases reach identical composition ($K < M$).
We can thus determine $K$ by clustering all $M$ phases based on the similarity of their final composition; see Methods.
This allows us to identify the phase count $K$ for a given interaction matrix~$\chi_{ij}$ and a given initial composition~$\p[n]_i$ of the phases.

We aim to characterize the distribution of the number~$K$ of distinct phases of a particular interaction matrix~$\chi_{ij}$ for the typical cellular situation where concentrations fluctuate widely.
In particular, the initial composition of the $M$ phases depends on the details of nucleation~\cite{Xu2014,Shimobayashi2021}.
To capture this, we sample initial compositions uniformly over all allowed volume fractions; see Appendix~\ref{sec:numerical_method} and Supporting \figref{fig:appendix_initial_composition}.
This ensemble defines a distribution~$P(K)$, which characterizes the behavior of a particular interaction matrix~$\chi_{ij}$; see \figref{fig:simulations}E.
In the cellular context, $P(K)$ corresponds to the frequency with which $K$ different condensates form simultaneously.
While cells surely also control compositions of these condensates, controlling their number is a more fundamental requirement, e.g., to prevent formation of aberrant condensates.

\subsection*{Random interactions do not lead to reliable phase counts}

\begin{figure*}[t]
	\centering	
	\includegraphics[width=17.8cm]{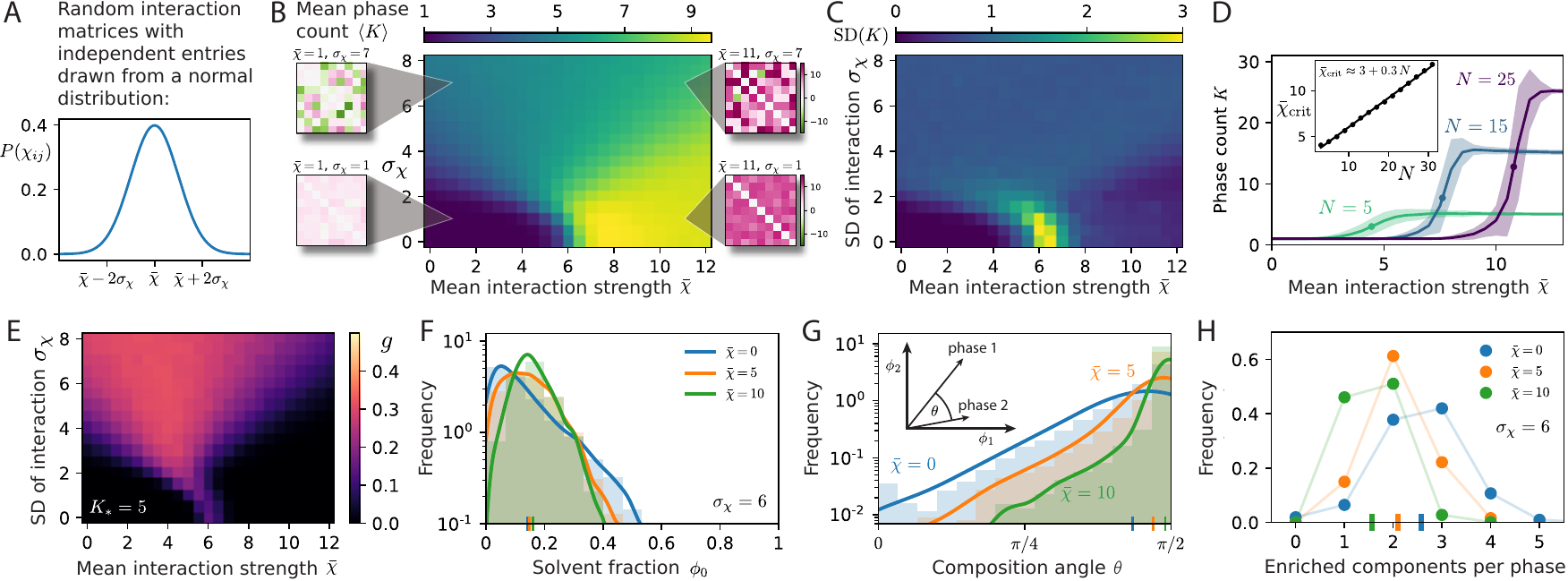}
	\caption{Random interaction matrices cannot target specific phase counts~$K$ reliably. 
	(A) Distribution $P(\chi_{ij})$ of the entries of the interaction matrix.
	(B, C) $\mean{K}$ and $\STD(K)$ as a function of the mean $\bar\chi$ and standard deviation $\sigma_\chi$ of the distribution for the interactions~$\chi_{ij}$ for $N=9$ components (insets show example matrices).
	(D)  $\mean{K} \pm \STD(K)$ as a function of $\bar\chi$ at $\sigma_\chi=1$ for $N=5, 15, 25$.
	The dot indicates the demixing transition point~$\bar\chi_\mathrm{crit}$.
	Inset: $\bar\chi_\mathrm{crit}$ as a function of the component count~$N$ (Line: $\bar\chi_\mathrm{crit} = 3 + 0.3 \, N$).
	(E) Performance~$g$ associated with data from panel (A) for $N=9$, $K_* = 5$, and $w=1$. %
	(F, G) Distribution of the solvent fraction~$\phi_0$ (F) and composition angles~$\theta$ (G) shown as histograms and using kernel density estimation (lines) for $N=9$,  $\sigma_\chi=6$, and several $\bar\chi$.
	(H) Distribution of the number of components enriched in phases for $N=9$,  $\sigma_\chi=6$, and several $\bar\chi$.
	(B--H) Averages are over $10^4$ realizations and distribution means are indicated as vertical bars on the horizontal axes.
	}
	\label{fig:random_matrices}
\end{figure*}

To gain intuition for the behavior of the multicomponent system, we first consider random interaction matrices~$\chi_{ij}$.
To compare with the literature~\cite{Sear2003,Jacobs2013,Jacobs2017,Shrinivas2021,Jacobs2021}, we draw entries independently from a normal distribution with mean $\bar\chi$ and variance $\sigma_\chi^2$; see \figref{fig:random_matrices}A.
For each parameter pair $(\bar\chi, \sigma^2_\chi)$, we investigate $10^4$ realizations of $\chi_{ij}$ and initial compositions and summarize the resulting distribution $P(K)$ by its mean and standard deviation.
\figref{fig:random_matrices}B shows that only a single phase forms when interactions are generally weak (low $\bar\chi$ and $\sigma_\chi$), consistent with an ideal solution where entropy favors mixing.
When interactions are increased without strong variations (larger $\bar\chi$, low $\sigma_\chi$), a demixing transition happens at $\bar\chi \approx \bar\chi_\mathrm{crit}$, and $K\approx N + 1$ phases are typical at large $\bar\chi$.
Here, all components segregate from each other and form separate phases, each enriched in a single component.
\figref{fig:random_matrices}C shows that the width of the phase count distribution, $\STD(K) = \mean{(K - \mean{K})^2}^{1/2}$, is largest in the transition zone, indicating that the actually observed~$K$ strongly depends on the chosen interaction matrix and initial composition.
The critical value~$\bar\chi_\mathrm{crit}$, where the demixing transition takes places, increases with the component count~$N$ (see \figref{fig:random_matrices}D), which confirms a trend that was observed in earlier work \cite{Jacobs2017, Sear2003}.
\figref{fig:random_matrices}B also shows that the width of the transition zone is generally broader for larger~$\sigma_\chi$, consistent with the fact that interactions are more variable.
Interestingly, the statistics of the phase count $K$ become independent of $\bar\chi$ for large variations~$\sigma_\chi$.
In this case, we observe $K \approx N/2$, which was previously conjectured for $\bar\chi= 0$~\cite{Shrinivas2021}.
Taken together, our simplified dynamics are consistent with known results for random matrices. %

The results shown in \figref{fig:random_matrices}B indicate that random interactions of $N$ components can lead to approximately $1$, $N/2$, and $N$ phases in large regions of the parameter space, while other values require fine-tuning.
Even if it is possible to find parameters $\bar\chi$ and $\sigma_\chi$ that on average lead to a desired phase count~$K_*$, it will not always be reached since the actual distribution of the number of phases, $P(K)$, possesses a significant width; see \figref{fig:random_matrices}C.
To quantify how well the system reaches a target phase count~$K_*$, we define the performance
\begin{equation}
	\label{eqn:performance}
	g = \sum_{K=1}^{K_\mathrm{max}} P(K) \exp\!\left[
		- \frac{(K-K_*)^2}{2w^2}
	\right]
	\;,
\end{equation}
which is constructed such that $0 < g \le 1$ and $g=1$ if and only if all initial conditions lead to $K_*$ phases.
Here, $w$ controls how strongly deviations from the target~$K_*$ are punished; see \figref{fig:simulations}E.
\figref{fig:random_matrices}E shows that the maximal performance of the random ensemble is $g\approx 0.3$, even though the choice $K_*=5$ is close to $N/2$, so large $\sigma_\chi$ leads to $\mean{K} \approx K_*$.
Indeed, supplementary Fig. S2 shows that the random ensemble performs even worse for other targets~$K_*$.
Taken together, it is thus not sufficient to vary the two parameters $\bar\chi$ and $\sigma_\chi$ of the random interactions to obtain a particular phase count reliably.

Equilibrium phases resulting from random interactions also show strong composition variations.
For instance, the solvent fraction~$\phi_0$ varies between $0$ and $0.5$ (\figref{fig:random_matrices}F). %
We quantify differences of phase compositions using the composition angle
$\theta_{nm} = \arccos(\vec\phi_n . \vec\phi_m / | \vec\phi_n | \cdot | \vec\phi_m |)$,
which is simply the angle between the composition vectors $\vec\phi_n = (\phi^{(n)}_1, \ldots, \phi^{(n)}_N)$ of  two phases $n$ and $m$~\cite{Shrinivas2021}; see inset of \figref{fig:random_matrices}G.
Note that $\theta$ is zero when phases have identical composition (but not necessarily the same total concentration) while $\theta=\frac{\pi}{2}$ when compositions are orthogonal, i.e., when they have no components in common.
While compositions of initial phases are similar (see Supporting \figref{fig:appendix_initial_composition}B), they typically become very different after equilibration; see \figref{fig:random_matrices}G.
In particular, the mean difference increases with stronger repulsion (larger $\bar\chi$).
However, even for the strongest repulsion, there is significant overlap between phases, indicating that components are not cleanly sorted into distinct phases.
To quantify this, we count for each phase how many components have a fraction larger than $1.5$ times the average fraction.
The number of such enriched components is smaller for stronger interactions, although it varies widely; see \figref{fig:random_matrices}H.
Taken together, typical random interaction matrices cannot provide a reliable phase count~$K$, so some additional structure is required.

\subsection*{Naively structured interactions also do not lead to reliable phase counts}

\begin{figure*}[t]
	\centering	
	\includegraphics[width=17.8cm]{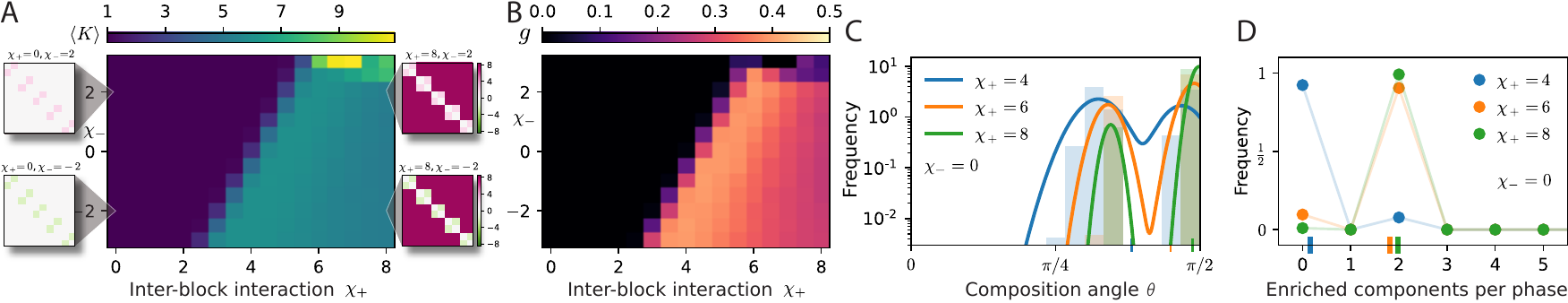}
	\caption{Interaction matrices with block structure cannot target specific phase counts~$K$ reliably. 
	(A, B) Mean phase count $\mean{K}$ and performance~$g$ as functions of the interaction $\chi_+$ between different blocks and the interaction $\chi_-$ within blocks for $N=10$ components arranged in $K_*=5$ equal blocks (insets in A show example matrices).
	(C)~Distribution of composition angles~$\theta$ shown as histograms and using kernel density estimation (lines) for several $\chi_+$ at $\chi_-=0$. 
	(D) Distribution of the number of components enriched in phases for several $\chi_+$ at $\chi_-=0$. 
	(A--D) %
	Averages over $10^4$ initial compositions;
	Distribution means are indicated by vertical bars.
	}
	\label{fig:block_matrices}
\end{figure*}

To elucidate what structure in interaction matrices reliably leads to a desired phase count~$K_*$, we next  group the $N$ interacting components in $K_*$ clusters.
We impose a repulsive interaction~$\chi_+$ between components belonging to different clusters, while components within a cluster exhibit a weak interaction~$\chi_-$.
We expect that components in the same cluster co-segregate, so the system behaves as if it consisted of $K_*$ effective components that all repel each other with strength $\chi_+$.
Indeed, \figref{fig:block_matrices}A shows that demixing into many phases happens when $\chi_+$ is sufficiently large, while the intra-cluster interaction $\chi_-$ has a weaker effect.
Co-segregation even takes place when the intra-cluster interaction is slightly repulsive ($\chi_->0$).
However, while these designed matrices display expected behavior, they still have significant variations and the resulting performance is only marginally better than that of random matrices (\figref{fig:block_matrices}B). 
This is also visible in the distribution of the composition angles~$\theta$ shown in \figref{fig:block_matrices}C:
Even for strong repulsion (large $\chi_+$) there is a significant fraction of phases with similar composition ($\theta \approx \pi/4$), even though exactly two components are enriched in each phase (\figref{fig:block_matrices}D).
It seems as if weakly concentrated components, including the solvent, prevent reliable co-segregation of  clustered components.
We thus find that creating interaction matrices with desired behavior is not as straight-forward as we had hoped.

\subsection*{Evolutionarily optimized interactions lead to reliable phase counts}

Neither completely random nor fully structured interaction matrices are very realistic in biology since the interaction energies~$\chi_{ij}$ summarize complex interactions of proteins~\cite{Dignon2020}, which change continuously during evolution~\cite{Schuster2020}.
We thus next ask whether an evolutionary optimization of interaction matrices can reliably lead to mixtures with a particular target phase count~$K_*$.

\begin{figure*}
	\centering	
	\includegraphics[width=17.8cm]{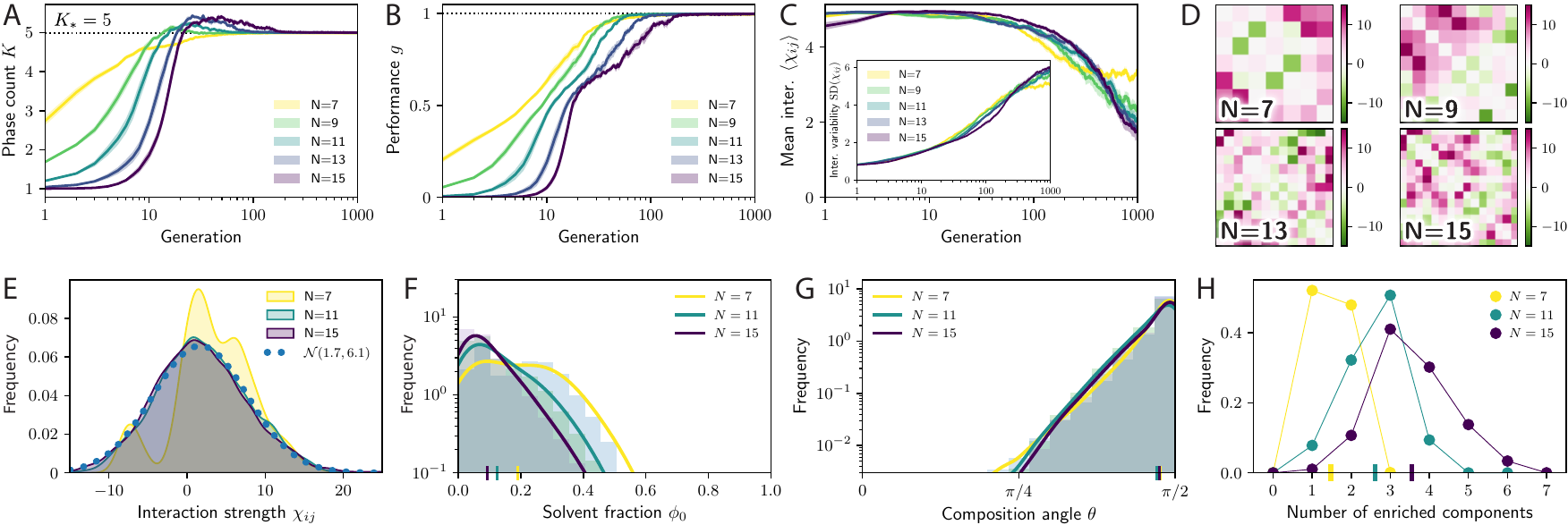}
	\caption{Evolutionary optimized interactions result in reliable phase counts~$K$.
	(A, B) Phase count $K$ and performance~$g$ as functions of generation for different number of components~$N$.
	The target phase count $K_*=5$ and the maximal performance $g=1$ are indicated by dotted lines.
	(C) Interaction strength~$\mean{\chi_{ij}}$ and associated standard deviation~$\mathrm{SD}(\chi_{ij})$ as a function of generation.
	(D) Examples for optimized interaction matrices for various $N$.
	Components have been clustered by similarity.
	(E) Distribution of interaction strengths in the final matrices. The blue dotted line indicates a normal distribution that best fits the case $N=15$.
	(F, G) Distribution of the solvent fraction~$\phi_0$ (F) and  composition angles~$\theta$ (G) shown as histograms and using kernel density estimation (lines) for several $N$. The corresponding means are indicated as vertical bars.
	(H) Distribution of the number of components enriched in phases for several $N$ with means indicated as vertical bars.
	(A--C) All curves show ensemble mean and associated standard error for $8$ repetitions.
	(A--H) Model parameters are $M=2N$, $K_*=5$, $w=1$, $\sigmaE=0.3$, and $\chiBound=5$.
	}
	\label{fig:evolution}
\end{figure*}

To mimic biology, we evolve an ensemble of individuals, characterized by interaction matrices~$\chi_{ij}$, over multiple generations.
We initialize a population of $32$ individuals with randomly chosen interaction matrices~$\chi_{ij}$ using $\bar\chi = \bar\chi_\mathrm{crit}$ and $\sigma_\chi=1$.
For each individual, we numerically determine $P(K)$ and the associated performance~$g$, see \Eqref{eqn:performance}, which will now play the role of a fitness.
In the selection step, we remove the $\unit[30]{\%}$ of the population with lowest performance, replacing them by randomly chosen high-performance individuals to maintain population size.
We then mutate the interactions~$\chi_{ij}$ of all individuals by adding independent random numbers from a normal distribution with zero mean and standard deviation~$\sigmaE$.
Repeating this procedure for many generations improves the performance of all individuals, so that they reliably reach the target~$K_*$.
However, we also noticed that this naive optimization results in very large interaction magnitudes (see Supporting \figref{fig:appendix_evolution_unconstraint}), which might be unrealistic.
To prevent such unphysical behavior, we additionally scale the interaction matrix~$\chi_{ij}$ by $\chiBound/\mean{|\chi_{ij}|}$ if its mean absolute value~$\mean{|\chi_{ij}|}$ exceeds the threshold $\chiBound$.
This limits the average interaction magnitude, $\mean{|\chi_{ij}|} \le \chiBound$, but the evolutionary optimization still discovers interaction matrices with a precise phase count (\figref{fig:evolution}A) and perfect performance (\figref{fig:evolution}B).
Optimized interaction matrices thus vastly outperform random matrices and allows targeting specific phase counts~$K_*$ despite strong fluctuations in initial composition.

The outstanding performance of evolved interaction matrices~$\chi_{ij}$ is surprising since we limited the interaction magnitude (\figref{fig:evolution}C) and use highly variable initial compositions.
What properties of $\chi_{ij}$ lead to the excellent performance~$g$?
Simply visualizing optimized interaction matrices (\figref{fig:evolution}E) does not reveal any obvious structure.
In any case, we showed in \figref{fig:block_matrices} that block matrices are not optimal, so any obvious clustering might actually be detrimental.
The distribution of the entires~$\chi_{ij}$ in optimized interaction matrices is very broad (\figref{fig:evolution}F), although its width is directly limited by our constraint of $\mean{|\chi_{ij}|}$.
For sufficiently large $N$, the distribution is well-described by a normal distribution (dotted blue line).
This is surprising, since unstructured random matrices chosen from such a normal distribution did not perform well (\figref{fig:random_matrices}).
The similarity to the random ensemble also shows in the distribution of the solvent fraction~$\phi_0$ (compare \figref{fig:evolution}F to \figref{fig:random_matrices}F) and the number of enriched components (compare \figref{fig:evolution}H to \figref{fig:random_matrices}H).
In contrast, the distribution of composition angles~$\theta$ is slightly different  (compare \figref{fig:evolution}G to \figref{fig:random_matrices}G).
However, $\mean{\theta}$ is larger for the random ensemble with large $\bar\chi$, implying more distinct phases.
Taken together,  optimized matrices share many similarities with random matrices, although minute differences apparently lead to a much improved performance.

The evolutionary optimization quickly discovered interaction matrices that lead to a reliable phase count and these matrices evolve continuously.
This begs the question whether this task is actually difficult;
how frequent are optimal matrices in the space of all matrices?
Our analysis of random matrices clearly showed that matrices must fulfill some basic requirements to have a phase count~$K$ close to the target~$K_*$.
In particular, the average magnitude of the entries~$\chi_{ij}$ and the associated standard deviation need to be chosen such that $\mean{K} \approx K_*$; see \figref{fig:random_matrices}B.
While we showed that the ensemble of random matrices with these properties does not work optimally (\figref{fig:random_matrices}E), individual matrices from the ensemble might perform well.
To quantify this, we determined the performance~$g$ for $64$ random matrices characterized by a particular choice of $\bar\chi$ and $\sigma_\chi$.
Supplementary \figref{fig:appendix_monte_carlo} shows that we easily discover matrices with high performance.
This implies that a large fraction of all matrices with suitable statistics, determined by $\mean{\chi_{ij}}$ and $\STD(\chi_{ij})$, performs optimally.

\subsection*{Performance of evolutionarily optimized interactions is robust}
We showed that interaction matrices leading to exactly~$K_*$ phases can be obtained through random trial-and-error or by evolutionary optimization.
This situation corresponds to maximizing the performance~$g$ in a fixed environment without any fluctuations beyond the initial composition.
However, biological systems constantly face additional fluctuations, both internally (e.g., changes of the component count) and externally (e.g., changing environment).
Such systems need to work not only in a particular case, but they need to be robust to these fluctuations, too.
To see how evolution of multicomponent phase separation fares in such challenging situations, we next  study the dynamics when interaction matrix~$\chi_{ij}$, the number~$N$ of components, or the target phase count~$K_*$ varies.

\begin{figure}
	\centering	
	\includegraphics[width=\columnwidth]{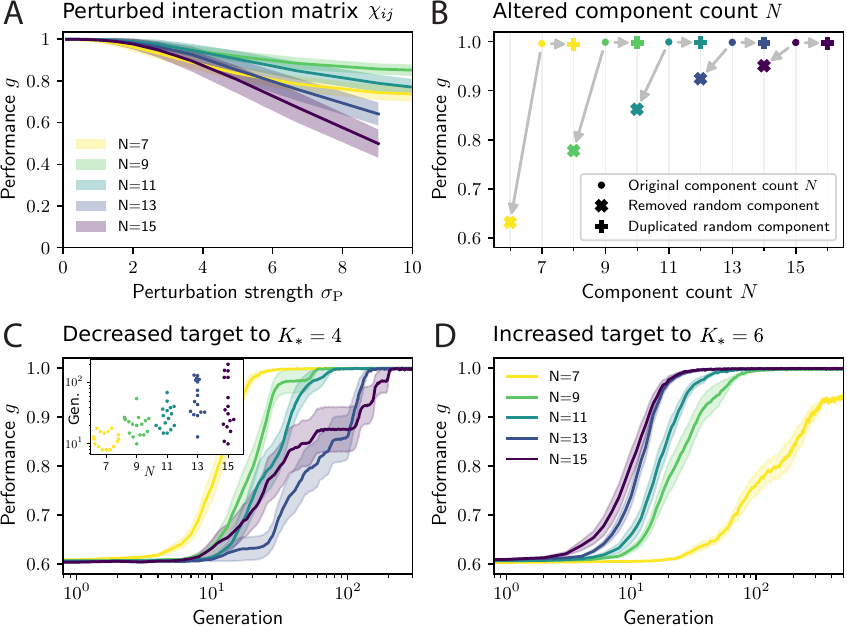}
	\caption{Mixtures with more components are typically more robust.
	(A)~Performance~$\mean{g} \pm \STD(g)$ of optimized interaction matrices where one row (and column) has been perturbed by normally distributed random numbers with standard deviation~$\sigmaP$ as a function of  $\sigmaP$ for several component counts~$N$ at $K_*=5$.
	(B)~Performance $g$ when a mixture optimized with~$N$ components loses (cross symbols) or duplicates (plus symbols) a component for various $N$ at $K_*=5$. SD is smaller than symbol size.
	(C, D)~Evolution of $\mean{g} \pm \SEM(g)$ as a function of generation when the target phase count is decreased from $K_*=5$ to $K_*=4$ (panel C) or increased from $K_*=5$ to $K_*=6$ (panel D) for several $N$.
	The inset in C shows the generation at which individual trajectories exceed a performance of $0.8$ for various~$N$.
	(A--D) Ensemble averages over the optimized matrices from Fig. 4 are shown.
	}
	\label{fig:robustness}
\end{figure}

We start by perturbing a single component in the evolutionarily optimized interaction matrices~$\chi_{ij}$ by choosing a random row (and column) to which we add uncorrelated random numbers from a normal distribution of vanishing mean and standard deviation~$\sigmaP$.
\figref{fig:robustness}A shows that the performance of optimized matrices is only weakly affected for $\sigmaP \lesssim 2$, while larger perturbations reduce the performance significantly.
Mixtures with more components are more sensitive to these perturbations, presumably because our procedure modifies the interaction between the chosen component and all other ones, so larger mixtures exhibit more perturbations.
Taken together, we find that optimized mixtures still form phases reliably even when interaction energies are perturbed by $\sim\kBT$.

We next test the robustness of the system against changes in component count~$N$ itself, which captures gene loss and duplication in real systems.
We quantify the effect of changing $N$ by measuring the performance~$g$ when one of the components of the optimized interactions matrices is removed or duplicated.
\figref{fig:robustness}B shows that removing a component reduces the performance substantially, although the reduction is smaller for larger $N$.
Conversely, duplicating a component has hardly any effect on performance.
Taken together, this suggests that using more components to form a fixed number of phases is more robust to internal fluctuations, like variations in component count.

We next consider external fluctuations of the environment.
Since we do not model the environment explicitly, we consider changes of the target phase count~$K_*$, assuming that the environment changes such that organisms need to form fewer or more phases.
Changing $K_*$ by one will necessarily reduce the performance from the optimal value $g\approx 1$ to $g = \exp(-\frac12w^{-2})\approx 0.6$; see \Eqref{eqn:performance}.
To see how well different systems adapt to new environments, we study how quickly the performance recovers under the evolutionary dynamics. %
\figref{fig:robustness}C shows that individuals quickly adjust to a lower target count, although the generation at which this happens varies widely; see inset.
This adaptation tends to be a bit slower for more components, presumably because more interactions have to be adjusted.
Conversely, adaptation to an increased target count~$K_*$ is easier for more components; see \figref{fig:robustness}D.
Note that the smallest system with $N=7$ does not succeed to meet the target~$K_*=6$ reliably due to the constraint on $\mean{|\chi_{ij}|}$. %
Taken together, this suggests that there is a larger flexibility in the phase composition at larger~$N$, which allows to quickly find an interaction matrix resulting in an additional phase.
Reducing the phase count is more complicated, likely because many interactions have to be adjusted.
In fact, there is a trade-off between robustly reaching a constant phase count~$K_*$ despite perturbations and using the same perturbations to flexibly adjust to new environments.

\section*{Discussion}
Understanding the equilibrium properties of biomolecules is crucial before we can tackle the more challenging problem of a living system.
We here proposed a novel method to study how the many interacting constituents of a cell spontaneously segregate into different phases.
This method recovers the demixing transition that was previously observed when many components exhibit random  interactions~\cite{Jacobs2017,Shrinivas2021}.
We also find a variable phase count for a given set of interactions, which is a signature of the complex phase diagrams~\cite{Mao2018}.
Beyond these limiting cases, our method can efficiently handle arbitrary interactions involving several tens of different components, thus increasing the range of systems that can be studied.

We use our method to optimize interaction matrices to yield a precise phase count.
These optimized interactions are also robust to perturbations and allow a fast adaption to an increased target phase count, particularly if many components are involved.
In contrast, forming fewer phases seems to be more challenging for larger mixtures, presumably because these mixtures are actually robust to perturbations.
It will be interesting to study this trade-off between robustness and evolvability in more detail in the future.

Optimal interaction matrices are surprisingly easy to discover and even random matrices have a high chance of yielding a robust number of phases, which is independent of the initial composition.
On the contrary, other random matrices from the same ensemble perform much worse. 
What are properties that separate the optimal matrices from generic ones?
Answering this question is directly relevant to biomolecular condensates, where hidden structures in intrinsically disordered regions might strongly affect the phase behavior of proteins~\cite{Moses2021}.
Our method can also be extended to describe more complex behavior of biomolecular condensates,
including response to external cues~\cite{Choi2020a,AdameArana2020}, active regulation~\cite{Hondele2020, Soeding2019, Kirschbaum2021}, and noise buffering~\cite{Klosin2020, Devirie2021}.
Ultimately, our predictions could be tested using engineered condensates~\cite{Bracha2019} and quantitative reconstitution~\cite{Currie2021}.
Beside these concrete applications for biomolecular condensates, our method might also answer more fundamental questions about evolving systems:
How can a cell exhibit robust functions while its proteins evolve~\cite{Laan2015,Diepeveen2018,Brauns2020a}?
We hope that our abstract model will illuminate the fundamental problem of how variable microscopic interactions lead to robust collective properties.

\begin{acknowledgments}
We thank Evan Spruijt for  a critical review of the manuscript and helpful discussions.
D.Z. acknowledges funding by the Max Planck Society.
L.L. gratefully acknowledges funding from the European Research Council (ERC) under the European Union’s Horizon 2020 research and innovation programme (Grant agreement No. 758132 and funding from the Netherlands Organization for Scientific Research (Nederlandse Organisatie voor Wetenschappelijk Onderzoek; NWO) through a VIDI grant (016.Vidi.171.060).
\end{acknowledgments}

\appendix
%%%%%%% SI

\setcounter{figure}{0}
    \renewcommand{\thefigure}{S\arabic{figure}}

\section{Derivation of simple dynamical system}
\label{sec:appendix_dynamical_system}
We here give a detailed derivation of the simple dynamical system that we use in the main text to determine equilibrium states of the multicomponent liquid.
In particular, we demonstrate below that the stationary states of this dynamical system correspond to actual equilibrium solutions.

\subsection{Equilibrium conditions of the physical system}
The conditions for an equilibrium between two phases $n$ and $m$ are given by \Eqref{eqn:equilibrium} in the main text.
The non-dimensional chemical potentials and pressures are
\begin{subequations}
\begin{align}
	\nm[n]_i &= 
		\ln(\p[n]_i) - \ln(\phiS[n]) + \sum_{j=1}^N \chi_{ij} \p[n]_j
\\
	\nP[n] &=
		- \ln(\phiS[n]) + \sum_{i,j=1}^N \frac{\chi_{ij}}{2} \p[n]_i\p[n]_j
	\;.
	\label{eqn:nondim_mu_P}
\end{align}
\end{subequations}
In general, the multiphase system is specified by the phase volumes~$\V[n]$ together with either all particle numbers, $\N[n]_i$, or all volume fractions~$\p[n]_i= \nu \N[n]_i/\V[n]$.
Here, $\V[n]$ and $\N[n]_i$ are extensive quantities that grow with system size, while the fractions~$\p[n]_i$ are the intensive quantities that appear in \Eqref{eqn:equilibrium} and \Eqref{eqn:nondim_mu_P}.

\subsection{Dynamics of the full physical system}
To determine the equilibrium states satisfying \Eqsref{eqn:equilibrium}, we first discuss the dynamics of the physical system, specified by the intensive quantities~$\p[n]_i$ together with the extensive volumes~$\V[n]$ of all phases.
We can then express the rate of change of the free energy $F=\sum_{n=1}^M \V[n] f(\p[n]_i)$ as
\begin{align}
	\partial_t F &= -\sum_{n=1}^M P^{(n)} \partial_t \V[n] 
		+ \sum_{i=1}^N \sum_{n=1}^M \m[n]_i \partial_t \N[n]_i
	\;.
	\label{eqn:free_energy_change}
\end{align}
Here, we consider the spontaneous relaxation to equilibrium, implying dynamics that decrease $F$ continuously, $\partial_t F \le 0$.
Using linear non-equilibrium thermodynamics~\cite{Julicher2018}, one can show that the exchange of volume is driven by pressure differences and differences in chemical potential imply changes in particle numbers,
\begin{subequations}
\label{eqn:linear_thermodynamics}
\begin{align}
	\partial_t \V[n] &= \nu \sum_m  \alpha^{(nm)}(\nP[n] - \nP[m])
\\
	\partial_t \N[n]_i &=  \sum_m \beta^{(nm)}_{i}(\nm[m]_i - \nm[n]_i)
	\;,
\end{align}
\end{subequations}
where the kinetic coefficients $\alpha^{(nm)}$ and $\beta^{(nm)}_{i}$ need to be symmetric, $\alpha^{(nm)} = \alpha^{(mn)}$ and $\beta^{(nm)}_{i} = \beta^{(mn)}_{i}$.
We can use this together with \Eqref{eqn:free_energy_change} to show explicitly that the free energy cannot increase under these conditions,
\begin{align}
	\frac{\partial_t F}{\kBT} &= 
		- \sum_{n,m=1}^M \frac{\alpha^{(nm)}}{2}\left(\nP[n] - \nP[m]\right)^2
\notag\\&\quad
		- \sum_{i=1}^N \sum_{n,m=1}^M \frac{\beta^{(nm)}_{i}}{2}\left(\nm[m]_i - \nm[n]_i\right)^2
	\;.
\end{align}
This implies that the following three statements about the system's state are all equivalent: (i) $F$ is at a stationary point, $\partial_t F=0$; (ii) The dynamics given in \Eqref{eqn:linear_thermodynamics} are at a stationary point; (iii) The system fulfills the equilibrium conditions \Eqref{eqn:equilibrium} and the volume and particle number constraints.
We can thus determine solutions to \Eqref{eqn:equilibrium} using the dynamics given in \Eqref{eqn:linear_thermodynamics} to relax initial conditions to a stationary state.

A concrete implementation of the relaxation dynamics requires sensible choices for the kinetic coefficients.
We use $\alpha^{(nm)}=k$ and $\beta^{(nm)}_{i}=k\p[n]_i\p[m]_i$, where we introduced the relaxation rate $k$, which defines the non-dimensional time $\hat t = kt$.
Using
$
	\partial_t \p[n]_i = [
		\nu \partial_t \N[n]_i - \p[n]_i \partial_t \V[n]
	] / \V[n]
$,
the dynamics of the system read
\begin{subequations}
\label{eqn:odes_full}
\begin{align}
	\partialT \V[n] &= \nu \sum_{m=1}^M \left[\nP[n] - \nP[m]\right]
	\label{eqn:odes_full_volume}
\\
	\partialT \p[n]_i &= \frac{\nu\p[n]_i}{\V[n]}
		\sum_{m=1}^M  \left[
			\p[m]_i \bigl(\nm[m]_i - \nm[n]_i\bigr)
			+ \bigl( \nP[m] - \nP[n]\bigr)
		\right]
	\label{eqn:odes_full_phi}
	\;.
\end{align}
\end{subequations}
These equations define an initial value problem, which relaxes an initial configuration, $\V[n]$ and $\p[n]_i$, toward equilibrium.

\subsection{Simplified dynamical system}
The physical dynamics defined by \Eqref{eqn:odes_full} involve both the intensive fractions~$\p[n]_i$ and the extensive volumes~$\V[n]$.
However, the equilibrium conditions (\ref{eqn:equilibrium}) are local statements about the coexistence of phases and thus do not involve extensive quantities.
To solve only the coexistence problem, we now seek a dynamical system that only involves intensive variables.
Inspecting \Eqref{eqn:odes_full_phi}, we see that the volumes~$\V[n]$ only affect the rate at which the fractions~$\p[n]_i$ change but not the direction of change.
Consequently, we obtain qualitatively similar dynamics by removing the pre-factor, which results in \Eqref{eqn:odes}.
Clearly, equilibrium states, which fulfill \Eqref{eqn:equilibrium}, are stationary states of \Eqref{eqn:odes}.
We next demonstrate that the $NM$ stationary state conditions of \Eqref{eqn:odes} also imply the equilibrium conditions~\Eqref{eqn:equilibrium}.
To do this, we introduce the deviations
\begin{align}
	\x[n]_i &= \nm[n]_i - \bar\mu_i
&
	\y[n] &= \nP[n] - \bar P
\end{align}
from the means $\bar\phi_i = M^{-1}\sum_n \p[n]_i$,  $\bar\mu_i = M^{-1}\sum_n \nm[n]_i$, and $\bar P = M^{-1} \sum_n\nP[n]$.
Using this, we can give the equilibrium conditions as $(N+1)M$ conditions,
\begin{align}
	\label{eqn:equilibrium_deviation}
	\x[n]_i &= 0
&
	 \y[n]&=0
& \forall \, n, i
	\;,
\end{align}
with $N+1$ constraints, $\sum_n \x[n]_i = \sum_n \y[n] = 0$,
resulting in $(N+1)(M-1)$ independent conditions.
In contrast, we have $NM$ conditions for the stationary state,
\begin{align}
	\label{eqn:stationary_states_deviation}
	0 &=  \frac{1}{M} \sum_{m=1}^M \p[m]_i \x[m]_i
	 	 - \bar\phi_i \x[n]_i
		- \y[n]
	&& \forall \, n, i
	\;.
\end{align}
Summing \Eqref{eqn:stationary_states_deviation} over $n$, we find $0= \sum_m \p[m]_i \x[m]_i$, which implies
\begin{align}
	0 &= \bar\phi_i \x[n]_i  + \y[n]
	&& \forall \, n, i \;.
\end{align}
Note that these are only $N(M-1)$ independent equations since $\sum_n \x[n]_i = \sum_n \y[n]=0$.
There are $N$ additional equations that follow from \Eqref{eqn:stationary_states_deviation}, resulting in $NM$ independent equations describing the stationary state,
\begin{subequations}
\begin{align}
	0 &= \bar\phi_i \x[n]_i + \y[n] \qquad \forall \, i \le N, n \le M - 1
\\
	0 &= \sum_{n=1}^M \p[n]_i  \x[n]_i \qquad \forall \, i
	\;.
\end{align}
\end{subequations}
To show that these are equivalent to the equilibrium conditions (\ref{eqn:equilibrium_deviation}), we express them as a linear system, 
\begin{align}
	\label{eqn:linear_system}
	0 = \mathcal A . X
	\;,
\end{align}
where
\begin{widetext}
\begin{align}
	X = (\x[1]_1, \ldots, \x[1]_N, \x[2]_1, \ldots, \x[2]_N, \ldots, \x[M-1]_1, \ldots, \x[M-1]_N, \y[1], \y[2], \ldots \y[M-1])
\end{align}
has $(M-1)(N+1)$ entries, and
\begin{align}
	\mathcal A &= \left(\begin{array}{@{}c|c|c|c||c@{}}
      \begin{matrix}
      \bar\phi_1 & 0 & \cdots & 0 \\
      0 & \bar\phi_2 & \cdots & 0 \\
      \vdots & \vdots & \ddots & \vdots \\
      0 & 0 & \cdots & \bar\phi_N
      \end{matrix}
      &  %
      \bigzero
      & \cdots 
      &  %
      \bigzero
      &
      \begin{matrix}
     1 & 0 & \cdots &0\\
      1 & 0 & \cdots &0\\
      \vdots & \vdots & \ddots& \vdots\\
      1 & 0 & \cdots &0\\
      \end{matrix}
      \\
\hline %
      \bigzero
      &  %
      \begin{matrix}
      \bar\phi_1 & 0 & \cdots & 0\\
      0 & \bar\phi_2 & \cdots & 0 \\
      \vdots & \vdots & \ddots & \vdots \\
      0 & 0 & \cdots & \bar\phi_N
      \end{matrix}
      & \cdots 
      &  %
      \bigzero
      &
      \begin{matrix}
     0 & 1 & \cdots & 0\\
      0 & 1 & \cdots & 0\\
      \vdots & \vdots & \ddots & \vdots \\
      0 & 1 & \cdots &0\\
      \end{matrix}
      \\
\hline
   \vdots & \vdots & \ddots & \vdots & \vdots \\
\hline %
      \bigzero 
      &  %
      \bigzero 
      & \cdots 
      &  %
      \begin{matrix}
      \bar\phi_1 & 0 & \cdots & 0 \\
      0 & \bar\phi_2 & \cdots & 0  \\
      \vdots & \vdots & \ddots & \vdots \\
      0 & 0 & \cdots & \bar\phi_N
      \end{matrix}
      &
      \begin{matrix}
     0 & 0 & \cdots & 1\\
      0 & 0 & \cdots & 1\\
      \vdots & \vdots & \ddots & \vdots \\
      0 & 0 & \cdots &1\\
      \end{matrix}
      \\
\hline\hline %
      \begin{matrix}
      \p[1]_1 & 0 & \cdots & 0 \\
      0 & \p[1]_2 & \cdots & 0 \\
      \vdots & \vdots & \ddots & \vdots \\
      0 & 0 & \cdots & \p[1]_N
      \end{matrix}
      &  %
      \begin{matrix}
      \p[2]_1 &  0 & \cdots & 0 \\
      0 & \p[2]_2 & \cdots & 0 \\
      \vdots & \vdots & \ddots & \vdots \\
      0 & 0 & \cdots & \p[2]_N
      \end{matrix}
      & \cdots 
      &  %
      \begin{matrix}
      \p[M-1]_1 & 0 & \cdots & 0  \\
      0 & \p[M-1]_2 & \cdots & 0 \\
      \vdots & \vdots & \ddots & \vdots  \\
      0 & 0 & \cdots & \p[M-1]_N
      \end{matrix}
      &
      \bigzero
      \\
\end{array}\right)
\end{align}
\end{widetext}
is a matrix of $M \times M$ blocks where each block has the dimension $N \times N$, except in the last column, where the blocks have dimension $N \times (M-1)$.
The block size can thus be summarized as
\begin{align}
	\begin{pmatrix}
	N \times N & \cdots & N \times N & N \times (M-1) \\
	\vdots & \ddots & \vdots & \vdots \\
	N \times N & \cdots & N \times N & N \times (M-1) \\
	\end{pmatrix}
	\;.
\end{align}
Clearly, this matrix has $MN$ rows and $(M-1)(N+1)$ columns, so it is a square matrix if and only if $M = N+1$.

To show that the stationary state conditions given in \Eqref{eqn:stationary_states_deviation} imply the equilibrium conditions \Eqref{eqn:equilibrium_deviation}, we need to show that the linear system given in \Eqref{eqn:linear_system} only has the trivial solution $X=0$.
If $\mathcal A$ is a square matrix, this amounts to showing that its determinant is non-zero.
We will show below that the non-square case can be treated by investigating the largest square sub matrix~$\mathcal A_\mathrm{sq}$, which is built by dropping the last rows or columns.
Defining the relevant dimension~$d=\min(M-1,N)$, the determinant of this matrix reads
\begin{align}
	\det(\mathcal A_\mathrm{sq})
	&= 
	(-1)^d 
	\det(\mathcal P)
	\prod_{i=1}^d \bar\phi_i^{M-2}
	\prod_{j=d+1}^{N} \bar\phi_j^{M-1}
	\;,
\end{align}
where we defined the square composition matrix
\begin{align}
	\mathcal P &= 
	\begin{pmatrix}
	\p[1]_1 & \p[2]_1 & \cdots & \p[d]_1 \\
	\p[1]_2 & \p[2]_2 & \cdots & \p[d]_2 \\
	\vdots & \vdots & \ddots & \vdots \\
	\p[1]_d & \p[2]_d & \cdots & \p[d]_d \\
	\end{pmatrix}
	\;.
\end{align}
Note that the determinant of $\mathcal A_\mathrm{sq}$ only vanishes if $\det(\mathcal P)=0$, since $\bar\phi_i>0$.
In the following, we analyze the solution space for the three relevant dimensional cases.

\subsubsection{Balanced case of a square matrix}
In the balance case, $M=N+1=d+1$, we have $\mathcal A=\mathcal A_\mathrm{sq}$ and $\mathcal P$ describes the full composition.
The linear system $0=\mathcal A.X$ could have non-trivial solutions if the determinant of $\mathcal P$ vanishes.
Generally, $\det(\mathcal P)$ vanishes if rows or columns are linearly dependent.
This is for instance the case when two phases have identical composition, $\p[n]_i = \p[m]_i$ for two phases $n\neq m$ and all components $i$.
However, in this case the conditions $\nm[n]_i = \nm[m]_i$ and $\nP[n] = \nP[m]$ are trivially fulfilled, so we can always ignore identical phases and instead focus on phases with distinct composition.
The determinant also vanishes when two species have identical composition in all phases, $\p[n]_i = \p[n]_j$ for some $i \neq j$ and all $n$.
This is only possible if they behave identically, $\chi_{ik} = \chi_{jk}$ for all $k$, in which case we again find $\nm[n]_i = \nm[n]_j$ and $\nP[n] = \nP[m]$.
In this case, the two components can basically be treated as one and thus also ignored.
Generally, the determinant also vanishes when a row (column) can be written as a linear sum of the other rows (columns).
While we could not identify a mathematical statement that this is impossible, we never observed such a case and will thus not discuss it further.
Taken together, we conclude that the stationary state conditions \Eqref{eqn:stationary_states_deviation} imply the equilibrium conditions~\Eqref{eqn:equilibrium_deviation} if $M=N+1$.

\subsubsection{Overdetermined case of few phases}
In this case, $\mathcal A$ has more rows than columns, so there are more stationary state conditions than equilibrium conditions.
This implies $NM > (N+1)(M-1)$, $M<N+1$,  and $d=M-1$. %
By solving the square sub-problem involving $\mathcal A_\mathrm{sq}$, we find $\x[n]_i=\y[n]=0$.
In doing so, we ignored the extra conditions
\begin{align}
	0 &= \sum_{n=1}^M \p[n]_i  \x[n]_i \qquad \text{for} \quad i = M, \ldots, N
	\;,
\end{align}
but these are trivially fulfilled.
Taken together, this shows that the overdetermined case implies  $\x[n]_i=\y[n]=0$.

\subsubsection{Underdetermined case of too many phases}
In this case, $\mathcal A$ has more columns than rows, $NM < (N+1)(M-1)$, $M>N+1$, and $d=N$.
This implies that there are not sufficient stationary state conditions to immediately conclude that the equilibrium conditions hold.
However, the variables $\x[n]_i$ and $\y[n]$ are not truly independent since they are both functions of the composition $\p[n]_i$; see \Eqref{eqn:nondim_mu_P}.
Combining these conditions, we find 
\begin{align}
	\sum_{i=1}^N \p[n]_i \nm[n]_i - 2 \nP[n] = \sum_{i=1}^N \p[n]_i \ln\p[n]_i + (1 + \phiS[n])\ln\phiS[n]
\end{align}
for all $n$, demonstrating a linear relationship between $\nm[n]_i$ and $\nP[n]$ that depends non-linearly on $\p[n]_i$.
Since these conditions must also hold, it is plausible that only $NM$ of the $(N+1)(M-1)$ unknowns of the linear system (\ref{eqn:linear_system}) are independent.
Assuming this is the case, we can solve the linear subsystem of $NM$ dimensions, whose determinant is given in \Eqref{eqn:nondim_mu_P}.
Similarly to the cases discussed above, this system has only the trivial solutions, implying that $\x[n]_i=\y[n]=0$ for the first $NM$ unknowns.
This shows that most equilibrium conditions follow from the stationary state conditions in the underdetermined case and suggests that the stationary state system does not possess any solutions that violate the equilibrium conditions.
Indeed, we never observed any numerical solutions of the stationary state conditions~(\ref{eqn:stationary_states_deviation}) that violated the equilibrium conditions~(\ref{eqn:equilibrium_deviation}).

Note that the underdetermined system is related to Gibbs' phase rule, which implies that at most $N+1$ phases can have different composition~\cite{Gibbs1876}.
Consequently, the number of undetermined variables we encountered here is exactly the minimal number of phases that needs to have a composition equivalent to other phases.
The additional constraints introduced by the fact that $\x[n]_i$ and $\y[n]$ all depend on $\p[n]_i$ thus reflect the thermodynamic stability discussed by Gibbs.

\section{Numerical solution method}
\label{sec:numerical_method}
We solve \Eqref{eqn:odes} using an explicit scheme with adaptive time stepping; see source code~\cite{sourcecode}. %
Since the simulation typically converges exponentially, we conclude that a stationary state has been reached when all $\partial_{\hat t} \p[n]_i < 10^{-4}$.
For each choice of $\chi_{ij}$, we run $64$ simulations with random initial conditions to estimate the distribution of the phase count~$K$, which is the minimal number of points $\vec{x}_m$ so that $\underset{m}{\min}\bigl(\| \vec{x}_m - \vec\phi^{(n)} \| \bigr) \le 10^{-2}$ for all phases $n$.

\begin{figure*}
\centering
\includegraphics[width=\textwidth]{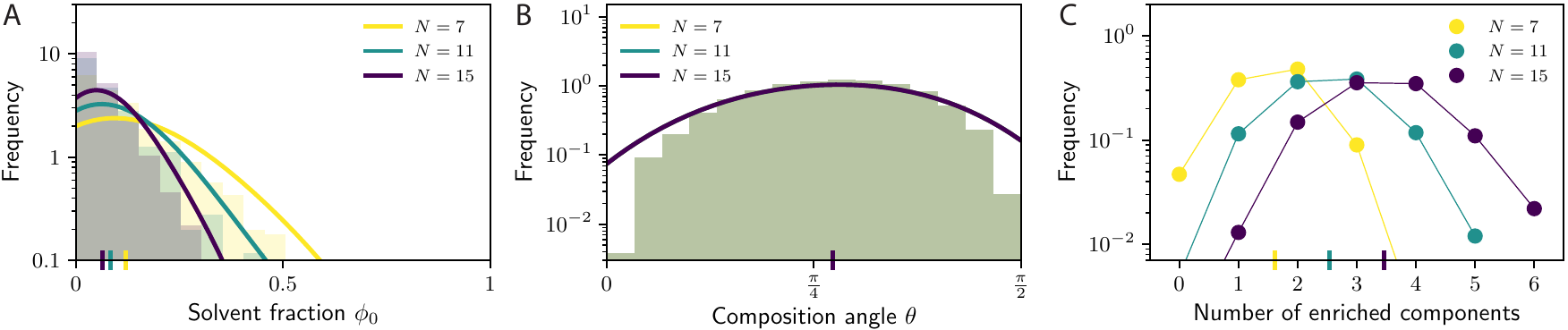}
\caption{The uniform initial compositions exhibit a large variability.
(A) Histograms and kernel density estimates of the solvent fraction~$\phi_0$ for various component counts~$N$. Since the composition is uniformly distributed, the fraction of each component follows the given distribution.
(B) Histograms and kernel density estimates of the composition angle~$\theta$ for~$N$, suggesting that this distribution is independent of $N$.
(C) Distribution of the number of components enriched in phases for various $N$.
(A--C) The means of the distributions are indicated by vertical bars on the horizontal axes.
}
\label{fig:appendix_initial_composition}
\end{figure*}

We choose initial conditions  $\vec\phi = (\phi_1, \ldots, \phi_N)$ for a phase such that all admissible composition vector $(\phi_0, \phi_1, \ldots, \phi_N)$ including the solvent fraction~$\phi_0$ exhibit a uniform distribution.
Here, we obviously use $\phi_0 = 1 - \sum_{i=1}^N \phi_i$ and we ensure $\phi_i \ge 0$ for $i=0,1, \ldots, N$.
The condition that all fractions sum to one implies correlations between the fractions, but since the geometry of allowed fractions is a simplex in an $N$-dimensional space, we can determine the marginal distributions along each dimension and choose the fractions iteratively.
We use the conditional probabilities
\begin{subequations}
\label{eqn:conditional_probabilities}
\begin{align}
	P_1(\phi_1)  &= P(\phi_1;1, N)	
\\
	P_2(\phi_2 | \phi_1) &= P(\phi_2; 1 - \phi_1, N-1)
\\
	&\vdots
\notag\\
	P_i(\phi_i| \phi_1, \ldots, \phi_{i-1}) &= P\biggl(\phi_i; 1 - \sum_{j=1}^{i-1} \phi_j, N - i + 1\biggr)
\\
	&\vdots
\notag\\
	P_N(\phi_N | \phi_1, \ldots, \phi_{N-1}) &= \frac{1}{1 - \sum_{j=1}^{N-1} \phi_j}
\end{align}
\end{subequations}
where
\begin{align}
	P(\phi; f, n) &= n \, (f - \phi)^{n-1}f^{-n}
\end{align}
is a particular, scaled Beta distribution defined for $\phi \in [0, f]$.
These conditional probabilities given in \Eqsref{eqn:conditional_probabilities} allow us to draw random variates of $\vec\phi$, since we can sample from the one-dimensional distributions for $i=1,\ldots,N$ one after another.
Note that the joint probability distribution reads
\begin{align}
	P(\phi_1, \ldots, \phi_N) &= \prod_{i=1}^N P_i(\phi_i | \phi_1, \ldots, \phi_{i-1})
	=N!
\end{align}
and is thus constant, demonstrating that this is truly a uniform distribution.
\figref{fig:appendix_initial_composition} shows statistics of these initial compositions.

To demonstrate the procedure, we consider the example $N=2$, where we have $P_1(\phi_1) = 2 \, (1 - \phi_1)$ and $P_2(\phi_2 | \phi_1) = (1 - \phi_1)^{-1}$, implying $P(\phi_1, \phi_2) = P_1(\phi_1)P_2(\phi_2 | \phi_1) = 2$. The expectation values are
\begin{subequations}
\begin{align}
	\mean{\phi_0} &= \int_0^1 \diff \phi_1 \int_0^{1-\phi_1} \diff \phi_2 \,
		P_{12}(\phi_1, \phi_2) \, (1 - \phi_1 -\phi_2)= \frac13
\\
	\mean{\phi_1} &= \int_0^1 \diff \phi_1 \int_0^{1-\phi_1} \diff \phi_2 \,
		P_{12}(\phi_1, \phi_2) \, \phi_1= \frac13
\\
	\mean{\phi_2} &= \int_0^1 \diff \phi_1 \int_0^{1-\phi_1} \diff \phi_2 \,
		P_{12}(\phi_1, \phi_2) \, \phi_2 = \frac13
\end{align}
\end{subequations}
demonstrating that indeed all three components have identical average fraction.

\bibliographystyle{apsrev4-2}
\bibliography{bibliography}

%apsrev4-2.bst 2019-01-14 (MD) hand-edited version of apsrev4-1.bst
%Control: key (0)
%Control: author (72) initials jnrlst
%Control: editor formatted (1) identically to author
%Control: production of article title (-1) disabled
%Control: page (0) single
%Control: year (1) truncated
%Control: production of eprint (0) enabled
\begin{thebibliography}{58}%
\makeatletter
\providecommand \@ifxundefined [1]{%
 \@ifx{#1\undefined}
}%
\providecommand \@ifnum [1]{%
 \ifnum #1\expandafter \@firstoftwo
 \else \expandafter \@secondoftwo
 \fi
}%
\providecommand \@ifx [1]{%
 \ifx #1\expandafter \@firstoftwo
 \else \expandafter \@secondoftwo
 \fi
}%
\providecommand \natexlab [1]{#1}%
\providecommand \enquote  [1]{``#1''}%
\providecommand \bibnamefont  [1]{#1}%
\providecommand \bibfnamefont [1]{#1}%
\providecommand \citenamefont [1]{#1}%
\providecommand \href@noop [0]{\@secondoftwo}%
\providecommand \href [0]{\begingroup \@sanitize@url \@href}%
\providecommand \@href[1]{\@@startlink{#1}\@@href}%
\providecommand \@@href[1]{\endgroup#1\@@endlink}%
\providecommand \@sanitize@url [0]{\catcode `\\12\catcode `\$12\catcode
  `\&12\catcode `\#12\catcode `\^12\catcode `\_12\catcode `\%12\relax}%
\providecommand \@@startlink[1]{}%
\providecommand \@@endlink[0]{}%
\providecommand \url  [0]{\begingroup\@sanitize@url \@url }%
\providecommand \@url [1]{\endgroup\@href {#1}{\urlprefix }}%
\providecommand \urlprefix  [0]{URL }%
\providecommand \Eprint [0]{\href }%
\providecommand \doibase [0]{https://doi.org/}%
\providecommand \selectlanguage [0]{\@gobble}%
\providecommand \bibinfo  [0]{\@secondoftwo}%
\providecommand \bibfield  [0]{\@secondoftwo}%
\providecommand \translation [1]{[#1]}%
\providecommand \BibitemOpen [0]{}%
\providecommand \bibitemStop [0]{}%
\providecommand \bibitemNoStop [0]{.\EOS\space}%
\providecommand \EOS [0]{\spacefactor3000\relax}%
\providecommand \BibitemShut  [1]{\csname bibitem#1\endcsname}%
\let\auto@bib@innerbib\@empty
%</preamble>
\bibitem [{\citenamefont {Brangwynne}\ \emph {et~al.}(2009)\citenamefont
  {Brangwynne}, \citenamefont {Eckmann}, \citenamefont {Courson}, \citenamefont
  {Rybarska}, \citenamefont {Hoege}, \citenamefont {Gharakhani}, \citenamefont
  {J{\"u}licher},\ and\ \citenamefont {Hyman}}]{Brangwynne2009}%
  \BibitemOpen
  \bibfield  {author} {\bibinfo {author} {\bibfnamefont {C.~P.}\ \bibnamefont
  {Brangwynne}}, \bibinfo {author} {\bibfnamefont {C.~R.}\ \bibnamefont
  {Eckmann}}, \bibinfo {author} {\bibfnamefont {D.~S.}\ \bibnamefont
  {Courson}}, \bibinfo {author} {\bibfnamefont {A.}~\bibnamefont {Rybarska}},
  \bibinfo {author} {\bibfnamefont {C.}~\bibnamefont {Hoege}}, \bibinfo
  {author} {\bibfnamefont {J.}~\bibnamefont {Gharakhani}}, \bibinfo {author}
  {\bibfnamefont {F.}~\bibnamefont {J{\"u}licher}},\ and\ \bibinfo {author}
  {\bibfnamefont {A.~A.}\ \bibnamefont {Hyman}},\ }\href
  {https://doi.org/10.1126/science.1172046} {\bibfield  {journal} {\bibinfo
  {journal} {Science}\ }\textbf {\bibinfo {volume} {324}},\ \bibinfo {pages}
  {1729} (\bibinfo {year} {2009})}\BibitemShut {NoStop}%
\bibitem [{\citenamefont {Feric}\ \emph {et~al.}(2016)\citenamefont {Feric},
  \citenamefont {Vaidya}, \citenamefont {Harmon}, \citenamefont {Mitrea},
  \citenamefont {Zhu}, \citenamefont {Richardson}, \citenamefont {Kriwacki},
  \citenamefont {Pappu},\ and\ \citenamefont {Brangwynne}}]{Feric2016}%
  \BibitemOpen
  \bibfield  {author} {\bibinfo {author} {\bibfnamefont {M.}~\bibnamefont
  {Feric}}, \bibinfo {author} {\bibfnamefont {N.}~\bibnamefont {Vaidya}},
  \bibinfo {author} {\bibfnamefont {T.~S.}\ \bibnamefont {Harmon}}, \bibinfo
  {author} {\bibfnamefont {D.~M.}\ \bibnamefont {Mitrea}}, \bibinfo {author}
  {\bibfnamefont {L.}~\bibnamefont {Zhu}}, \bibinfo {author} {\bibfnamefont
  {T.~M.}\ \bibnamefont {Richardson}}, \bibinfo {author} {\bibfnamefont
  {R.~W.}\ \bibnamefont {Kriwacki}}, \bibinfo {author} {\bibfnamefont {R.~V.}\
  \bibnamefont {Pappu}},\ and\ \bibinfo {author} {\bibfnamefont {C.~P.}\
  \bibnamefont {Brangwynne}},\ }\href
  {https://doi.org/10.1016/j.cell.2016.04.047} {\bibfield  {journal} {\bibinfo
  {journal} {Cell}\ }\textbf {\bibinfo {volume} {165}},\ \bibinfo {pages}
  {1686} (\bibinfo {year} {2016})}\BibitemShut {NoStop}%
\bibitem [{\citenamefont {Banani}\ \emph {et~al.}(2017)\citenamefont {Banani},
  \citenamefont {Lee}, \citenamefont {Hyman},\ and\ \citenamefont
  {Rosen}}]{Banani2017}%
  \BibitemOpen
  \bibfield  {author} {\bibinfo {author} {\bibfnamefont {S.~F.}\ \bibnamefont
  {Banani}}, \bibinfo {author} {\bibfnamefont {H.~O.}\ \bibnamefont {Lee}},
  \bibinfo {author} {\bibfnamefont {A.~A.}\ \bibnamefont {Hyman}},\ and\
  \bibinfo {author} {\bibfnamefont {M.~K.}\ \bibnamefont {Rosen}},\ }\href
  {https://doi.org/10.1038/nrm.2017.7} {\bibfield  {journal} {\bibinfo
  {journal} {Nat. Rev. Mol. Cell Biol.}\ }\textbf {\bibinfo {volume} {18}},\
  \bibinfo {pages} {285} (\bibinfo {year} {2017})}\BibitemShut {NoStop}%
\bibitem [{\citenamefont {Azaldegui}\ \emph {et~al.}(2020)\citenamefont
  {Azaldegui}, \citenamefont {Vecchiarelli},\ and\ \citenamefont
  {Biteen}}]{Azaldegui2020}%
  \BibitemOpen
  \bibfield  {author} {\bibinfo {author} {\bibfnamefont {C.~A.}\ \bibnamefont
  {Azaldegui}}, \bibinfo {author} {\bibfnamefont {A.~G.}\ \bibnamefont
  {Vecchiarelli}},\ and\ \bibinfo {author} {\bibfnamefont {J.~S.}\ \bibnamefont
  {Biteen}},\ }\href
  {https://doi.org/https://doi.org/10.1016/j.bpj.2020.09.023} {\bibfield
  {journal} {\bibinfo  {journal} {Biophys. J.}\ }\textbf {\bibinfo {volume}
  {120}},\ \bibinfo {pages} {1123} (\bibinfo {year} {2020})}\BibitemShut
  {NoStop}%
\bibitem [{\citenamefont {Cohan}\ and\ \citenamefont
  {Pappu}(2020)}]{Cohan2020}%
  \BibitemOpen
  \bibfield  {author} {\bibinfo {author} {\bibfnamefont {M.~C.}\ \bibnamefont
  {Cohan}}\ and\ \bibinfo {author} {\bibfnamefont {R.~V.}\ \bibnamefont
  {Pappu}},\ }\href {https://doi.org/10.1016/j.tibs.2020.04.011} {\bibfield
  {journal} {\bibinfo  {journal} {Trends Biochem Sci}\ }\textbf {\bibinfo
  {volume} {45}},\ \bibinfo {pages} {668} (\bibinfo {year} {2020})}\BibitemShut
  {NoStop}%
\bibitem [{\citenamefont {Greening}\ and\ \citenamefont
  {Lithgow}(2020)}]{Greening2020}%
  \BibitemOpen
  \bibfield  {author} {\bibinfo {author} {\bibfnamefont {C.}~\bibnamefont
  {Greening}}\ and\ \bibinfo {author} {\bibfnamefont {T.}~\bibnamefont
  {Lithgow}},\ }\href {https://doi.org/10.1038/s41579-020-0413-0} {\bibfield
  {journal} {\bibinfo  {journal} {Nat. Rev. Microbiol.}\ }\textbf {\bibinfo
  {volume} {18}},\ \bibinfo {pages} {677} (\bibinfo {year} {2020})}\BibitemShut
  {NoStop}%
\bibitem [{\citenamefont {Emenecker}\ \emph {et~al.}(2021)\citenamefont
  {Emenecker}, \citenamefont {Holehouse},\ and\ \citenamefont
  {Strader}}]{Emenecker2021}%
  \BibitemOpen
  \bibfield  {author} {\bibinfo {author} {\bibfnamefont {R.~J.}\ \bibnamefont
  {Emenecker}}, \bibinfo {author} {\bibfnamefont {A.~S.}\ \bibnamefont
  {Holehouse}},\ and\ \bibinfo {author} {\bibfnamefont {L.~C.}\ \bibnamefont
  {Strader}},\ }\bibfield  {journal} {\bibinfo  {journal} {Annu. Rev. Plant
  Biol.}\ }\textbf {\bibinfo {volume} {72}},\ \href
  {https://doi.org/10.1146/annurev-arplant-081720-015238}
  {10.1146/annurev-arplant-081720-015238} (\bibinfo {year} {2021}),\ \bibinfo
  {note} {pMID: 33684296}\BibitemShut {NoStop}%
\bibitem [{\citenamefont {Kim}\ \emph {et~al.}(2021)\citenamefont {Kim},
  \citenamefont {Lee}, \citenamefont {Lee},\ and\ \citenamefont
  {Seo}}]{Kim2021}%
  \BibitemOpen
  \bibfield  {author} {\bibinfo {author} {\bibfnamefont {J.}~\bibnamefont
  {Kim}}, \bibinfo {author} {\bibfnamefont {H.}~\bibnamefont {Lee}}, \bibinfo
  {author} {\bibfnamefont {H.~G.}\ \bibnamefont {Lee}},\ and\ \bibinfo {author}
  {\bibfnamefont {P.~J.}\ \bibnamefont {Seo}},\ }\href
  {https://doi.org/https://doi.org/10.15252/embr.202051656} {\bibfield
  {journal} {\bibinfo  {journal} {EMBO reports}\ }\textbf {\bibinfo {volume}
  {22}},\ \bibinfo {pages} {e51656} (\bibinfo {year} {2021})}\BibitemShut
  {NoStop}%
\bibitem [{\citenamefont {Lafontaine}\ \emph {et~al.}(2020)\citenamefont
  {Lafontaine}, \citenamefont {Riback}, \citenamefont {Bascetin},\ and\
  \citenamefont {Brangwynne}}]{Lafontaine2020}%
  \BibitemOpen
  \bibfield  {author} {\bibinfo {author} {\bibfnamefont {D.~L.~J.}\
  \bibnamefont {Lafontaine}}, \bibinfo {author} {\bibfnamefont {J.~A.}\
  \bibnamefont {Riback}}, \bibinfo {author} {\bibfnamefont {R.}~\bibnamefont
  {Bascetin}},\ and\ \bibinfo {author} {\bibfnamefont {C.~P.}\ \bibnamefont
  {Brangwynne}},\ }\href {https://doi.org/10.1038/s41580-020-0272-6} {\bibfield
   {journal} {\bibinfo  {journal} {Nat. Rev. Mol. Cell Biol.}\ }\textbf
  {\bibinfo {volume} {22}},\ \bibinfo {pages} {165} (\bibinfo {year}
  {2020})}\BibitemShut {NoStop}%
\bibitem [{\citenamefont {Fei}\ \emph {et~al.}(2017)\citenamefont {Fei},
  \citenamefont {Jadaliha}, \citenamefont {Harmon}, \citenamefont {Li},
  \citenamefont {Hua}, \citenamefont {Hao}, \citenamefont {Holehouse},
  \citenamefont {Reyer}, \citenamefont {Sun}, \citenamefont {Freier},
  \citenamefont {Pappu}, \citenamefont {Prasanth},\ and\ \citenamefont
  {Ha}}]{Fei2017}%
  \BibitemOpen
  \bibfield  {author} {\bibinfo {author} {\bibfnamefont {J.}~\bibnamefont
  {Fei}}, \bibinfo {author} {\bibfnamefont {M.}~\bibnamefont {Jadaliha}},
  \bibinfo {author} {\bibfnamefont {T.~S.}\ \bibnamefont {Harmon}}, \bibinfo
  {author} {\bibfnamefont {I.~T.~S.}\ \bibnamefont {Li}}, \bibinfo {author}
  {\bibfnamefont {B.}~\bibnamefont {Hua}}, \bibinfo {author} {\bibfnamefont
  {Q.}~\bibnamefont {Hao}}, \bibinfo {author} {\bibfnamefont {A.~S.}\
  \bibnamefont {Holehouse}}, \bibinfo {author} {\bibfnamefont {M.}~\bibnamefont
  {Reyer}}, \bibinfo {author} {\bibfnamefont {Q.}~\bibnamefont {Sun}}, \bibinfo
  {author} {\bibfnamefont {S.~M.}\ \bibnamefont {Freier}}, \bibinfo {author}
  {\bibfnamefont {R.~V.}\ \bibnamefont {Pappu}}, \bibinfo {author}
  {\bibfnamefont {K.~V.}\ \bibnamefont {Prasanth}},\ and\ \bibinfo {author}
  {\bibfnamefont {T.}~\bibnamefont {Ha}},\ }\href
  {https://doi.org/10.1242/jcs.206854} {\bibfield  {journal} {\bibinfo
  {journal} {J. Cell Sci.}\ }\textbf {\bibinfo {volume} {130}},\ \bibinfo
  {pages} {4180} (\bibinfo {year} {2017})}\BibitemShut {NoStop}%
\bibitem [{\citenamefont {Fritsch}\ \emph {et~al.}(2021)\citenamefont
  {Fritsch}, \citenamefont {Diaz-Delgadillo}, \citenamefont {Adame-Arana},
  \citenamefont {Hoege}, \citenamefont {Mittasch}, \citenamefont {Kreysing},
  \citenamefont {Leaver}, \citenamefont {Hyman}, \citenamefont {J{\"u}licher},\
  and\ \citenamefont {Weber}}]{Fritsch2021}%
  \BibitemOpen
  \bibfield  {author} {\bibinfo {author} {\bibfnamefont {A.~W.}\ \bibnamefont
  {Fritsch}}, \bibinfo {author} {\bibfnamefont {A.~F.}\ \bibnamefont
  {Diaz-Delgadillo}}, \bibinfo {author} {\bibfnamefont {O.}~\bibnamefont
  {Adame-Arana}}, \bibinfo {author} {\bibfnamefont {C.}~\bibnamefont {Hoege}},
  \bibinfo {author} {\bibfnamefont {M.}~\bibnamefont {Mittasch}}, \bibinfo
  {author} {\bibfnamefont {M.}~\bibnamefont {Kreysing}}, \bibinfo {author}
  {\bibfnamefont {M.}~\bibnamefont {Leaver}}, \bibinfo {author} {\bibfnamefont
  {A.~A.}\ \bibnamefont {Hyman}}, \bibinfo {author} {\bibfnamefont
  {F.}~\bibnamefont {J{\"u}licher}},\ and\ \bibinfo {author} {\bibfnamefont
  {C.~A.}\ \bibnamefont {Weber}},\ }\bibfield  {journal} {\bibinfo  {journal}
  {Proc. Natl. Acad. Sci. USA}\ }\textbf {\bibinfo {volume} {118}},\ \href
  {https://doi.org/10.1073/pnas.2102772118} {10.1073/pnas.2102772118} (\bibinfo
  {year} {2021})\BibitemShut {NoStop}%
\bibitem [{\citenamefont {Dignon}\ \emph {et~al.}(2020)\citenamefont {Dignon},
  \citenamefont {Best},\ and\ \citenamefont {Mittal}}]{Dignon2020}%
  \BibitemOpen
  \bibfield  {author} {\bibinfo {author} {\bibfnamefont {G.~L.}\ \bibnamefont
  {Dignon}}, \bibinfo {author} {\bibfnamefont {R.~B.}\ \bibnamefont {Best}},\
  and\ \bibinfo {author} {\bibfnamefont {J.}~\bibnamefont {Mittal}},\ }\href
  {https://doi.org/10.1146/annurev-physchem-071819-113553} {\bibfield
  {journal} {\bibinfo  {journal} {Annu. Rev. Phys. Chem.}\ }\textbf {\bibinfo
  {volume} {71}},\ \bibinfo {pages} {53} (\bibinfo {year} {2020})}\BibitemShut
  {NoStop}%
\bibitem [{\citenamefont {Hardenberg}\ \emph {et~al.}(2020)\citenamefont
  {Hardenberg}, \citenamefont {Horvath}, \citenamefont {Ambrus}, \citenamefont
  {Fuxreiter},\ and\ \citenamefont {Vendruscolo}}]{Hardenberg2020}%
  \BibitemOpen
  \bibfield  {author} {\bibinfo {author} {\bibfnamefont {M.}~\bibnamefont
  {Hardenberg}}, \bibinfo {author} {\bibfnamefont {A.}~\bibnamefont {Horvath}},
  \bibinfo {author} {\bibfnamefont {V.}~\bibnamefont {Ambrus}}, \bibinfo
  {author} {\bibfnamefont {M.}~\bibnamefont {Fuxreiter}},\ and\ \bibinfo
  {author} {\bibfnamefont {M.}~\bibnamefont {Vendruscolo}},\ }\href
  {https://doi.org/10.1073/pnas.2007670117} {\bibfield  {journal} {\bibinfo
  {journal} {Proc. Natl. Acad. Sci. USA}\ }\textbf {\bibinfo {volume} {117}},\
  \bibinfo {pages} {33254} (\bibinfo {year} {2020})}\BibitemShut {NoStop}%
\bibitem [{\citenamefont {Adekunle}\ and\ \citenamefont
  {Hubstenberger}(2020)}]{Adekunle2020}%
  \BibitemOpen
  \bibfield  {author} {\bibinfo {author} {\bibfnamefont {D.~A.}\ \bibnamefont
  {Adekunle}}\ and\ \bibinfo {author} {\bibfnamefont {A.}~\bibnamefont
  {Hubstenberger}},\ }\href {https://doi.org/10.1042/ETLS20190187} {\bibfield
  {journal} {\bibinfo  {journal} {Emerg Top Life Sci}\ }\textbf {\bibinfo
  {volume} {4}},\ \bibinfo {pages} {265} (\bibinfo {year} {2020})}\BibitemShut
  {NoStop}%
\bibitem [{\citenamefont {Lyon}\ \emph {et~al.}(2020)\citenamefont {Lyon},
  \citenamefont {Peeples},\ and\ \citenamefont {Rosen}}]{Lyon2020}%
  \BibitemOpen
  \bibfield  {author} {\bibinfo {author} {\bibfnamefont {A.~S.}\ \bibnamefont
  {Lyon}}, \bibinfo {author} {\bibfnamefont {W.~B.}\ \bibnamefont {Peeples}},\
  and\ \bibinfo {author} {\bibfnamefont {M.~K.}\ \bibnamefont {Rosen}},\ }\href
  {https://doi.org/10.1038/s41580-020-00303-z} {\bibfield  {journal} {\bibinfo
  {journal} {Nat. Rev. Mol. Cell Biol.}\ }\textbf {\bibinfo {volume} {22}},\
  \bibinfo {pages} {215} (\bibinfo {year} {2020})}\BibitemShut {NoStop}%
\bibitem [{\citenamefont {Jin}\ \emph {et~al.}(2021)\citenamefont {Jin},
  \citenamefont {Lee}, \citenamefont {Schaefer}, \citenamefont {Luo},
  \citenamefont {Wollman}, \citenamefont {Payne-Dwyer}, \citenamefont {Tian},
  \citenamefont {Zhang}, \citenamefont {Chen}, \citenamefont {Li},
  \citenamefont {McLeish}, \citenamefont {Leake},\ and\ \citenamefont
  {Bai}}]{Jin2021}%
  \BibitemOpen
  \bibfield  {author} {\bibinfo {author} {\bibfnamefont {X.}~\bibnamefont
  {Jin}}, \bibinfo {author} {\bibfnamefont {J.-E.}\ \bibnamefont {Lee}},
  \bibinfo {author} {\bibfnamefont {C.}~\bibnamefont {Schaefer}}, \bibinfo
  {author} {\bibfnamefont {X.}~\bibnamefont {Luo}}, \bibinfo {author}
  {\bibfnamefont {A.~J.~M.}\ \bibnamefont {Wollman}}, \bibinfo {author}
  {\bibfnamefont {A.~L.}\ \bibnamefont {Payne-Dwyer}}, \bibinfo {author}
  {\bibfnamefont {T.}~\bibnamefont {Tian}}, \bibinfo {author} {\bibfnamefont
  {X.}~\bibnamefont {Zhang}}, \bibinfo {author} {\bibfnamefont
  {X.}~\bibnamefont {Chen}}, \bibinfo {author} {\bibfnamefont {Y.}~\bibnamefont
  {Li}}, \bibinfo {author} {\bibfnamefont {T.~C.~B.}\ \bibnamefont {McLeish}},
  \bibinfo {author} {\bibfnamefont {M.~C.}\ \bibnamefont {Leake}},\ and\
  \bibinfo {author} {\bibfnamefont {F.}~\bibnamefont {Bai}},\ }\href
  {https://doi.org/10.1126/sciadv.abh2929} {\bibfield  {journal} {\bibinfo
  {journal} {Sci. Adv.}\ }\textbf {\bibinfo {volume} {7}},\ \bibinfo {pages}
  {eabh2929} (\bibinfo {year} {2021})}\BibitemShut {NoStop}%
\bibitem [{\citenamefont {Alberti}\ and\ \citenamefont
  {Dormann}(2019)}]{Alberti2019a}%
  \BibitemOpen
  \bibfield  {author} {\bibinfo {author} {\bibfnamefont {S.}~\bibnamefont
  {Alberti}}\ and\ \bibinfo {author} {\bibfnamefont {D.}~\bibnamefont
  {Dormann}},\ }\href {https://doi.org/10.1146/annurev-genet-112618-043527}
  {\bibfield  {journal} {\bibinfo  {journal} {Annu. Rev. Genet.}\ }\textbf
  {\bibinfo {volume} {53}},\ \bibinfo {pages} {171} (\bibinfo {year} {2019})},\
  \bibinfo {note} {pMID: 31430179}\BibitemShut {NoStop}%
\bibitem [{\citenamefont {Saar}\ \emph {et~al.}(2021)\citenamefont {Saar},
  \citenamefont {Morgunov}, \citenamefont {Qi}, \citenamefont {Arter},
  \citenamefont {Krainer}, \citenamefont {Lee},\ and\ \citenamefont
  {Knowles}}]{Saare2021}%
  \BibitemOpen
  \bibfield  {author} {\bibinfo {author} {\bibfnamefont {K.~L.}\ \bibnamefont
  {Saar}}, \bibinfo {author} {\bibfnamefont {A.~S.}\ \bibnamefont {Morgunov}},
  \bibinfo {author} {\bibfnamefont {R.}~\bibnamefont {Qi}}, \bibinfo {author}
  {\bibfnamefont {W.~E.}\ \bibnamefont {Arter}}, \bibinfo {author}
  {\bibfnamefont {G.}~\bibnamefont {Krainer}}, \bibinfo {author} {\bibfnamefont
  {A.~A.}\ \bibnamefont {Lee}},\ and\ \bibinfo {author} {\bibfnamefont
  {T.~P.~J.}\ \bibnamefont {Knowles}},\ }\bibfield  {journal} {\bibinfo
  {journal} {Proc. Natl. Acad. Sci. USA}\ }\textbf {\bibinfo {volume} {118}},\
  \href {https://doi.org/10.1073/pnas.2019053118} {10.1073/pnas.2019053118}
  (\bibinfo {year} {2021})\BibitemShut {NoStop}%
\bibitem [{\citenamefont {Choi}\ \emph {et~al.}(2020)\citenamefont {Choi},
  \citenamefont {Holehouse},\ and\ \citenamefont {Pappu}}]{Choi2020a}%
  \BibitemOpen
  \bibfield  {author} {\bibinfo {author} {\bibfnamefont {J.-M.}\ \bibnamefont
  {Choi}}, \bibinfo {author} {\bibfnamefont {A.~S.}\ \bibnamefont
  {Holehouse}},\ and\ \bibinfo {author} {\bibfnamefont {R.~V.}\ \bibnamefont
  {Pappu}},\ }\href {https://doi.org/10.1146/annurev-biophys-121219-081629}
  {\bibfield  {journal} {\bibinfo  {journal} {Annu. Rev. Biophys.}\ }\textbf
  {\bibinfo {volume} {49}},\ \bibinfo {pages} {107} (\bibinfo {year}
  {2020})}\BibitemShut {NoStop}%
\bibitem [{\citenamefont {Harmon}\ \emph {et~al.}(2017)\citenamefont {Harmon},
  \citenamefont {Holehouse}, \citenamefont {Rosen},\ and\ \citenamefont
  {Pappu}}]{Harmon2017}%
  \BibitemOpen
  \bibfield  {author} {\bibinfo {author} {\bibfnamefont {T.~S.}\ \bibnamefont
  {Harmon}}, \bibinfo {author} {\bibfnamefont {A.~S.}\ \bibnamefont
  {Holehouse}}, \bibinfo {author} {\bibfnamefont {M.~K.}\ \bibnamefont
  {Rosen}},\ and\ \bibinfo {author} {\bibfnamefont {R.~V.}\ \bibnamefont
  {Pappu}},\ }\bibfield  {journal} {\bibinfo  {journal} {Elife}\ }\textbf
  {\bibinfo {volume} {6}},\ \href {https://doi.org/10.7554/eLife.30294}
  {10.7554/eLife.30294} (\bibinfo {year} {2017})\BibitemShut {NoStop}%
\bibitem [{\citenamefont {Lin}\ \emph {et~al.}(2018)\citenamefont {Lin},
  \citenamefont {Forman-Kay},\ and\ \citenamefont {Chan}}]{Lin2018}%
  \BibitemOpen
  \bibfield  {author} {\bibinfo {author} {\bibfnamefont {Y.-H.}\ \bibnamefont
  {Lin}}, \bibinfo {author} {\bibfnamefont {J.~D.}\ \bibnamefont
  {Forman-Kay}},\ and\ \bibinfo {author} {\bibfnamefont {H.~S.}\ \bibnamefont
  {Chan}},\ }\href {https://doi.org/10.1021/acs.biochem.8b00058} {\bibfield
  {journal} {\bibinfo  {journal} {Biochemistry}\ }\textbf {\bibinfo {volume}
  {57}},\ \bibinfo {pages} {2499} (\bibinfo {year} {2018})}\BibitemShut
  {NoStop}%
\bibitem [{\citenamefont {Schuster}\ \emph {et~al.}(2020)\citenamefont
  {Schuster}, \citenamefont {Dignon}, \citenamefont {Tang}, \citenamefont
  {Kelley}, \citenamefont {Ranganath}, \citenamefont {Jahnke}, \citenamefont
  {Simpkins}, \citenamefont {Regy}, \citenamefont {Hammer}, \citenamefont
  {Good} \emph {et~al.}}]{Schuster2020}%
  \BibitemOpen
  \bibfield  {author} {\bibinfo {author} {\bibfnamefont {B.~S.}\ \bibnamefont
  {Schuster}}, \bibinfo {author} {\bibfnamefont {G.~L.}\ \bibnamefont
  {Dignon}}, \bibinfo {author} {\bibfnamefont {W.~S.}\ \bibnamefont {Tang}},
  \bibinfo {author} {\bibfnamefont {F.~M.}\ \bibnamefont {Kelley}}, \bibinfo
  {author} {\bibfnamefont {A.~K.}\ \bibnamefont {Ranganath}}, \bibinfo {author}
  {\bibfnamefont {C.~N.}\ \bibnamefont {Jahnke}}, \bibinfo {author}
  {\bibfnamefont {A.~G.}\ \bibnamefont {Simpkins}}, \bibinfo {author}
  {\bibfnamefont {R.~M.}\ \bibnamefont {Regy}}, \bibinfo {author}
  {\bibfnamefont {D.~A.}\ \bibnamefont {Hammer}}, \bibinfo {author}
  {\bibfnamefont {M.~C.}\ \bibnamefont {Good}}, \emph {et~al.},\ }\href@noop {}
  {\bibfield  {journal} {\bibinfo  {journal} {Proc. Natl. Acad. Sci. USA}\
  }\textbf {\bibinfo {volume} {117}},\ \bibinfo {pages} {11421} (\bibinfo
  {year} {2020})}\BibitemShut {NoStop}%
\bibitem [{\citenamefont {Bremer}\ \emph {et~al.}(2021)\citenamefont {Bremer},
  \citenamefont {Farag}, \citenamefont {Borcherds}, \citenamefont {Peran},
  \citenamefont {Martin}, \citenamefont {Pappu},\ and\ \citenamefont
  {Mittag}}]{Bremer2021}%
  \BibitemOpen
  \bibfield  {author} {\bibinfo {author} {\bibfnamefont {A.}~\bibnamefont
  {Bremer}}, \bibinfo {author} {\bibfnamefont {M.}~\bibnamefont {Farag}},
  \bibinfo {author} {\bibfnamefont {W.~M.}\ \bibnamefont {Borcherds}}, \bibinfo
  {author} {\bibfnamefont {I.}~\bibnamefont {Peran}}, \bibinfo {author}
  {\bibfnamefont {E.~W.}\ \bibnamefont {Martin}}, \bibinfo {author}
  {\bibfnamefont {R.~V.}\ \bibnamefont {Pappu}},\ and\ \bibinfo {author}
  {\bibfnamefont {T.}~\bibnamefont {Mittag}},\ }\bibfield  {journal} {\bibinfo
  {journal} {Nat. Chem.}\ }\href {https://doi.org/10.1038/s41557-021-00840-w}
  {10.1038/s41557-021-00840-w} (\bibinfo {year} {2021})\BibitemShut {NoStop}%
\bibitem [{\citenamefont {Hyman}\ \emph {et~al.}(2014)\citenamefont {Hyman},
  \citenamefont {Weber},\ and\ \citenamefont {J{\"u}licher}}]{Hyman2014}%
  \BibitemOpen
  \bibfield  {author} {\bibinfo {author} {\bibfnamefont {A.~A.}\ \bibnamefont
  {Hyman}}, \bibinfo {author} {\bibfnamefont {C.~A.}\ \bibnamefont {Weber}},\
  and\ \bibinfo {author} {\bibfnamefont {F.}~\bibnamefont {J{\"u}licher}},\
  }\href@noop {} {\bibfield  {journal} {\bibinfo  {journal} {Annu. Rev. Cell
  Dev. Biol.}\ }\textbf {\bibinfo {volume} {30}},\ \bibinfo {pages} {39}
  (\bibinfo {year} {2014})}\BibitemShut {NoStop}%
\bibitem [{\citenamefont {Brangwynne}\ \emph {et~al.}(2015)\citenamefont
  {Brangwynne}, \citenamefont {Tompa},\ and\ \citenamefont
  {Pappu}}]{Brangwynne2015}%
  \BibitemOpen
  \bibfield  {author} {\bibinfo {author} {\bibfnamefont {C.~P.}\ \bibnamefont
  {Brangwynne}}, \bibinfo {author} {\bibfnamefont {P.}~\bibnamefont {Tompa}},\
  and\ \bibinfo {author} {\bibfnamefont {R.~V.}\ \bibnamefont {Pappu}},\
  }\href@noop {} {\bibfield  {journal} {\bibinfo  {journal} {Nat. Phys.}\
  }\textbf {\bibinfo {volume} {11}},\ \bibinfo {pages} {899} (\bibinfo {year}
  {2015})}\BibitemShut {NoStop}%
\bibitem [{\citenamefont {Berry}\ \emph {et~al.}(2018)\citenamefont {Berry},
  \citenamefont {Brangwynne},\ and\ \citenamefont {Haataja}}]{Berry2018}%
  \BibitemOpen
  \bibfield  {author} {\bibinfo {author} {\bibfnamefont {J.}~\bibnamefont
  {Berry}}, \bibinfo {author} {\bibfnamefont {C.}~\bibnamefont {Brangwynne}},\
  and\ \bibinfo {author} {\bibfnamefont {M.~P.}\ \bibnamefont {Haataja}},\
  }\href {https://doi.org/10.1088/1361-6633/aaa61e} {\bibfield  {journal}
  {\bibinfo  {journal} {Rep. Prog. Phys.}\ }\textbf {\bibinfo {volume} {81}},\
  \bibinfo {pages} {046601} (\bibinfo {year} {2018})}\BibitemShut {NoStop}%
\bibitem [{\citenamefont {Weber}\ \emph {et~al.}(2019)\citenamefont {Weber},
  \citenamefont {Zwicker}, \citenamefont {J{\"u}licher},\ and\ \citenamefont
  {Lee}}]{Weber2019}%
  \BibitemOpen
  \bibfield  {author} {\bibinfo {author} {\bibfnamefont {C.~A.}\ \bibnamefont
  {Weber}}, \bibinfo {author} {\bibfnamefont {D.}~\bibnamefont {Zwicker}},
  \bibinfo {author} {\bibfnamefont {F.}~\bibnamefont {J{\"u}licher}},\ and\
  \bibinfo {author} {\bibfnamefont {C.~F.}\ \bibnamefont {Lee}},\ }\href
  {https://iopscience.iop.org/article/10.1088/1361-6633/ab052b} {\bibfield
  {journal} {\bibinfo  {journal} {Rep. Prog. Phys.}\ }\textbf {\bibinfo
  {volume} {82}},\ \bibinfo {pages} {064601} (\bibinfo {year}
  {2019})}\BibitemShut {NoStop}%
\bibitem [{\citenamefont {Riback}\ \emph {et~al.}(2020)\citenamefont {Riback},
  \citenamefont {Zhu}, \citenamefont {Ferrolino}, \citenamefont {Tolbert},
  \citenamefont {Mitrea}, \citenamefont {Sanders}, \citenamefont {Wei},
  \citenamefont {Kriwacki},\ and\ \citenamefont {Brangwynne}}]{Riback2020}%
  \BibitemOpen
  \bibfield  {author} {\bibinfo {author} {\bibfnamefont {J.~A.}\ \bibnamefont
  {Riback}}, \bibinfo {author} {\bibfnamefont {L.}~\bibnamefont {Zhu}},
  \bibinfo {author} {\bibfnamefont {M.~C.}\ \bibnamefont {Ferrolino}}, \bibinfo
  {author} {\bibfnamefont {M.}~\bibnamefont {Tolbert}}, \bibinfo {author}
  {\bibfnamefont {D.~M.}\ \bibnamefont {Mitrea}}, \bibinfo {author}
  {\bibfnamefont {D.~W.}\ \bibnamefont {Sanders}}, \bibinfo {author}
  {\bibfnamefont {M.-T.}\ \bibnamefont {Wei}}, \bibinfo {author} {\bibfnamefont
  {R.~W.}\ \bibnamefont {Kriwacki}},\ and\ \bibinfo {author} {\bibfnamefont
  {C.~P.}\ \bibnamefont {Brangwynne}},\ }\href
  {https://doi.org/10.1038/s41586-020-2256-2} {\bibfield  {journal} {\bibinfo
  {journal} {Nature}\ }\textbf {\bibinfo {volume} {581}},\ \bibinfo {pages}
  {209} (\bibinfo {year} {2020})}\BibitemShut {NoStop}%
\bibitem [{\citenamefont {Mao}\ \emph {et~al.}(2018)\citenamefont {Mao},
  \citenamefont {Kuldinow}, \citenamefont {Haataja},\ and\ \citenamefont
  {Kosmrlj}}]{Mao2018}%
  \BibitemOpen
  \bibfield  {author} {\bibinfo {author} {\bibfnamefont {S.}~\bibnamefont
  {Mao}}, \bibinfo {author} {\bibfnamefont {D.}~\bibnamefont {Kuldinow}},
  \bibinfo {author} {\bibfnamefont {M.}~\bibnamefont {Haataja}},\ and\ \bibinfo
  {author} {\bibfnamefont {A.}~\bibnamefont {Kosmrlj}},\ }\href@noop {}
  {\bibfield  {journal} {\bibinfo  {journal} {Soft Matter}\ }\textbf {\bibinfo
  {volume} {15}},\ \bibinfo {pages} {1297} (\bibinfo {year}
  {2018})}\BibitemShut {NoStop}%
\bibitem [{\citenamefont {Mao}\ \emph {et~al.}(2020)\citenamefont {Mao},
  \citenamefont {Chakraverti-Wuerthwein}, \citenamefont {Gaudio},\ and\
  \citenamefont {Ko\ifmmode~\check{s}\else \v{s}\fi{}mrlj}}]{Mao2020}%
  \BibitemOpen
  \bibfield  {author} {\bibinfo {author} {\bibfnamefont {S.}~\bibnamefont
  {Mao}}, \bibinfo {author} {\bibfnamefont {M.~S.}\ \bibnamefont
  {Chakraverti-Wuerthwein}}, \bibinfo {author} {\bibfnamefont {H.}~\bibnamefont
  {Gaudio}},\ and\ \bibinfo {author} {\bibfnamefont {A.}~\bibnamefont
  {Ko\ifmmode~\check{s}\else \v{s}\fi{}mrlj}},\ }\href
  {https://doi.org/10.1103/PhysRevLett.125.218003} {\bibfield  {journal}
  {\bibinfo  {journal} {Phys. Rev. Lett.}\ }\textbf {\bibinfo {volume} {125}},\
  \bibinfo {pages} {218003} (\bibinfo {year} {2020})}\BibitemShut {NoStop}%
\bibitem [{\citenamefont {Leung}\ \emph {et~al.}(2003)\citenamefont {Leung},
  \citenamefont {Andersen}, \citenamefont {Mann},\ and\ \citenamefont
  {Lamond}}]{Leung2003}%
  \BibitemOpen
  \bibfield  {author} {\bibinfo {author} {\bibfnamefont {A.~K.~L.}\
  \bibnamefont {Leung}}, \bibinfo {author} {\bibfnamefont {J.~S.}\ \bibnamefont
  {Andersen}}, \bibinfo {author} {\bibfnamefont {M.}~\bibnamefont {Mann}},\
  and\ \bibinfo {author} {\bibfnamefont {A.~I.}\ \bibnamefont {Lamond}},\
  }\href {https://doi.org/10.1042/bj20031169} {\bibfield  {journal} {\bibinfo
  {journal} {Biochem. J}\ }\textbf {\bibinfo {volume} {376}},\ \bibinfo {pages}
  {553} (\bibinfo {year} {2003})}\BibitemShut {NoStop}%
\bibitem [{\citenamefont {Updike}\ and\ \citenamefont
  {Strome}(2009)}]{Updike2009}%
  \BibitemOpen
  \bibfield  {author} {\bibinfo {author} {\bibfnamefont {D.~L.}\ \bibnamefont
  {Updike}}\ and\ \bibinfo {author} {\bibfnamefont {S.}~\bibnamefont
  {Strome}},\ }\href {https://doi.org/10.1534/genetics.109.110171} {\bibfield
  {journal} {\bibinfo  {journal} {Genetics}\ }\textbf {\bibinfo {volume}
  {183}},\ \bibinfo {pages} {1397} (\bibinfo {year} {2009})}\BibitemShut
  {NoStop}%
\bibitem [{\citenamefont {Currie}\ and\ \citenamefont
  {Rosen}(2021)}]{Currie2021}%
  \BibitemOpen
  \bibfield  {author} {\bibinfo {author} {\bibfnamefont {S.~L.}\ \bibnamefont
  {Currie}}\ and\ \bibinfo {author} {\bibfnamefont {M.~K.}\ \bibnamefont
  {Rosen}},\ }\href {https://doi.org/10.1261/rna.079008.121} {\bibfield
  {journal} {\bibinfo  {journal} {RNA}\ }\textbf {\bibinfo {volume} {28}},\
  \bibinfo {pages} {27} (\bibinfo {year} {2021})}\BibitemShut {NoStop}%
\bibitem [{\citenamefont {Banani}\ \emph {et~al.}(2016)\citenamefont {Banani},
  \citenamefont {Rice}, \citenamefont {Peeples}, \citenamefont {Lin},
  \citenamefont {Jain}, \citenamefont {Parker},\ and\ \citenamefont
  {Rosen}}]{Banani2016}%
  \BibitemOpen
  \bibfield  {author} {\bibinfo {author} {\bibfnamefont {S.~F.}\ \bibnamefont
  {Banani}}, \bibinfo {author} {\bibfnamefont {A.~M.}\ \bibnamefont {Rice}},
  \bibinfo {author} {\bibfnamefont {W.~B.}\ \bibnamefont {Peeples}}, \bibinfo
  {author} {\bibfnamefont {Y.}~\bibnamefont {Lin}}, \bibinfo {author}
  {\bibfnamefont {S.}~\bibnamefont {Jain}}, \bibinfo {author} {\bibfnamefont
  {R.}~\bibnamefont {Parker}},\ and\ \bibinfo {author} {\bibfnamefont {M.~K.}\
  \bibnamefont {Rosen}},\ }\href
  {https://doi.org/https://doi.org/10.1016/j.cell.2016.06.010} {\bibfield
  {journal} {\bibinfo  {journal} {Cell}\ }\textbf {\bibinfo {volume} {166}},\
  \bibinfo {pages} {651} (\bibinfo {year} {2016})}\BibitemShut {NoStop}%
\bibitem [{\citenamefont {Zhou}\ and\ \citenamefont {Xie}(2021)}]{Zhou2021a}%
  \BibitemOpen
  \bibfield  {author} {\bibinfo {author} {\bibfnamefont {S.}~\bibnamefont
  {Zhou}}\ and\ \bibinfo {author} {\bibfnamefont {Y.~M.}\ \bibnamefont {Xie}},\
  }\href {https://doi.org/https://doi.org/10.1016/j.ijmecsci.2021.106349}
  {\bibfield  {journal} {\bibinfo  {journal} {International Journal of
  Mechanical Sciences}\ }\textbf {\bibinfo {volume} {198}},\ \bibinfo {pages}
  {106349} (\bibinfo {year} {2021})}\BibitemShut {NoStop}%
\bibitem [{\citenamefont {Shrinivas}\ and\ \citenamefont
  {Brenner}(2021)}]{Shrinivas2021}%
  \BibitemOpen
  \bibfield  {author} {\bibinfo {author} {\bibfnamefont {K.}~\bibnamefont
  {Shrinivas}}\ and\ \bibinfo {author} {\bibfnamefont {M.~P.}\ \bibnamefont
  {Brenner}},\ }\bibfield  {journal} {\bibinfo  {journal} {Proc. Natl. Acad.
  Sci. USA}\ }\textbf {\bibinfo {volume} {118}},\ \href
  {https://doi.org/10.1073/pnas.2108551118} {10.1073/pnas.2108551118} (\bibinfo
  {year} {2021})\BibitemShut {NoStop}%
\bibitem [{\citenamefont {Sear}\ and\ \citenamefont {Cuesta}(2003)}]{Sear2003}%
  \BibitemOpen
  \bibfield  {author} {\bibinfo {author} {\bibfnamefont {R.~P.}\ \bibnamefont
  {Sear}}\ and\ \bibinfo {author} {\bibfnamefont {J.}~\bibnamefont {Cuesta}},\
  }\bibfield  {booktitle} {\emph {\bibinfo {booktitle} {Physical Review
  Letters}},\ }\href
  {http://eutils.ncbi.nlm.nih.gov/entrez/eutils/elink.fcgi?dbfrom=pubmed\&id=14683134\&retmode=ref\&cmd=prlinks}
  {\bibfield  {journal} {\bibinfo  {journal} {Phys. Rev. Lett.}\ }\textbf
  {\bibinfo {volume} {91}},\ \bibinfo {pages} {245701} (\bibinfo {year}
  {2003})}\BibitemShut {NoStop}%
\bibitem [{\citenamefont {Jacobs}\ and\ \citenamefont
  {Frenkel}(2013)}]{Jacobs2013}%
  \BibitemOpen
  \bibfield  {author} {\bibinfo {author} {\bibfnamefont {W.~M.}\ \bibnamefont
  {Jacobs}}\ and\ \bibinfo {author} {\bibfnamefont {D.}~\bibnamefont
  {Frenkel}},\ }\bibfield  {booktitle} {\emph {\bibinfo {booktitle} {The
  Journal of Chemical Physics}},\ }\href {https://doi.org/10.1063/1.4812461}
  {\bibfield  {journal} {\bibinfo  {journal} {J. Chem. Phys.}\ }\textbf
  {\bibinfo {volume} {139}},\ \bibinfo {pages} {024108} (\bibinfo {year}
  {2013})}\BibitemShut {NoStop}%
\bibitem [{\citenamefont {Jacobs}\ and\ \citenamefont
  {Frenkel}(2017)}]{Jacobs2017}%
  \BibitemOpen
  \bibfield  {author} {\bibinfo {author} {\bibfnamefont {W.~M.}\ \bibnamefont
  {Jacobs}}\ and\ \bibinfo {author} {\bibfnamefont {D.}~\bibnamefont
  {Frenkel}},\ }\href {https://doi.org/10.1016/j.bpj.2016.10.043} {\bibfield
  {journal} {\bibinfo  {journal} {Biophys. J.}\ }\textbf {\bibinfo {volume}
  {112}},\ \bibinfo {pages} {683} (\bibinfo {year} {2017})}\BibitemShut
  {NoStop}%
\bibitem [{\citenamefont {Jacobs}(2021)}]{Jacobs2021}%
  \BibitemOpen
  \bibfield  {author} {\bibinfo {author} {\bibfnamefont {W.~M.}\ \bibnamefont
  {Jacobs}},\ }\href {https://doi.org/10.1103/PhysRevLett.126.258101}
  {\bibfield  {journal} {\bibinfo  {journal} {Phys. Rev. Lett.}\ }\textbf
  {\bibinfo {volume} {126}},\ \bibinfo {pages} {258101} (\bibinfo {year}
  {2021})}\BibitemShut {NoStop}%
\bibitem [{\citenamefont {Gibbs}(1876)}]{Gibbs1876}%
  \BibitemOpen
  \bibfield  {author} {\bibinfo {author} {\bibfnamefont {J.~W.}\ \bibnamefont
  {Gibbs}},\ }\bibfield  {booktitle} {\emph {\bibinfo {booktitle} {Transactions
  of the Connecticut Academy of Arts and Sciences}},\ }\href@noop {} {\bibfield
   {journal} {\bibinfo  {journal} {Trans. Conn. Acad. Arts Sci.}\ }\textbf
  {\bibinfo {volume} {3}},\ \bibinfo {pages} {1} (\bibinfo {year}
  {1876})}\BibitemShut {NoStop}%
\bibitem [{\citenamefont {Flory}(1942)}]{Flory1942}%
  \BibitemOpen
  \bibfield  {author} {\bibinfo {author} {\bibfnamefont {P.~I.}\ \bibnamefont
  {Flory}},\ }\bibfield  {booktitle} {\emph {\bibinfo {booktitle} {The Journal
  of Chemical Physics}},\ }\href@noop {} {\bibfield  {journal} {\bibinfo
  {journal} {J. Chem. Phys.}\ }\textbf {\bibinfo {volume} {10}},\ \bibinfo
  {pages} {51} (\bibinfo {year} {1942})}\BibitemShut {NoStop}%
\bibitem [{\citenamefont {Cahn}\ and\ \citenamefont
  {Hilliard}(1958)}]{Cahn1958}%
  \BibitemOpen
  \bibfield  {author} {\bibinfo {author} {\bibfnamefont {J.~W.}\ \bibnamefont
  {Cahn}}\ and\ \bibinfo {author} {\bibfnamefont {J.~E.}\ \bibnamefont
  {Hilliard}},\ }\href
  {http://apps.webofknowledge.com/InboundService.do?SID=V1McfEFn8CFjIn4a989\&product=WOS\&UT=A1958WA15300013\&SrcApp=Papers\&DestFail=http\%3A\%2F\%2Faccess.isiproducts.com\%2Fcustom\_images\%2Fwok5\_failed\_auth.html\&Init=Yes\&action=retrieve\&SrcAuth=mekentosj\&customersID=mekentosj\&mode=FullRecord}
  {\bibfield  {journal} {\bibinfo  {journal} {J. Chem. Phys.}\ }\textbf
  {\bibinfo {volume} {28}},\ \bibinfo {pages} {258} (\bibinfo {year}
  {1958})}\BibitemShut {NoStop}%
\bibitem [{\citenamefont {Xu}\ \emph {et~al.}(2014)\citenamefont {Xu},
  \citenamefont {Ting}, \citenamefont {Kusaka},\ and\ \citenamefont
  {Wang}}]{Xu2014}%
  \BibitemOpen
  \bibfield  {author} {\bibinfo {author} {\bibfnamefont {X.}~\bibnamefont
  {Xu}}, \bibinfo {author} {\bibfnamefont {C.~L.}\ \bibnamefont {Ting}},
  \bibinfo {author} {\bibfnamefont {I.}~\bibnamefont {Kusaka}},\ and\ \bibinfo
  {author} {\bibfnamefont {Z.-G.}\ \bibnamefont {Wang}},\ }\href
  {https://doi.org/10.1146/annurev-physchem-032511-143750} {\bibfield
  {journal} {\bibinfo  {journal} {Annu. Rev. Phys. Chem.}\ }\textbf {\bibinfo
  {volume} {65}},\ \bibinfo {pages} {449} (\bibinfo {year} {2014})},\ \bibinfo
  {note} {pMID: 24689799}\BibitemShut {NoStop}%
\bibitem [{\citenamefont {Shimobayashi}\ \emph {et~al.}(2021)\citenamefont
  {Shimobayashi}, \citenamefont {Ronceray}, \citenamefont {Sanders},
  \citenamefont {Haataja},\ and\ \citenamefont
  {Brangwynne}}]{Shimobayashi2021}%
  \BibitemOpen
  \bibfield  {author} {\bibinfo {author} {\bibfnamefont {S.~F.}\ \bibnamefont
  {Shimobayashi}}, \bibinfo {author} {\bibfnamefont {P.}~\bibnamefont
  {Ronceray}}, \bibinfo {author} {\bibfnamefont {D.~W.}\ \bibnamefont
  {Sanders}}, \bibinfo {author} {\bibfnamefont {M.~P.}\ \bibnamefont
  {Haataja}},\ and\ \bibinfo {author} {\bibfnamefont {C.~P.}\ \bibnamefont
  {Brangwynne}},\ }\href {https://doi.org/10.1038/s41586-021-03905-5}
  {\bibfield  {journal} {\bibinfo  {journal} {Nature}\ }\textbf {\bibinfo
  {volume} {599}},\ \bibinfo {pages} {503} (\bibinfo {year}
  {2021})}\BibitemShut {NoStop}%
\bibitem [{\citenamefont {Moses}\ \emph {et~al.}(2021)\citenamefont {Moses},
  \citenamefont {Guadalupe}, \citenamefont {Yu}, \citenamefont {Flores},
  \citenamefont {Perez}, \citenamefont {McAnelly}, \citenamefont {Shamoon},
  \citenamefont {Cuevas-Zepeda}, \citenamefont {Merg}, \citenamefont {Martin},
  \citenamefont {Holehouse},\ and\ \citenamefont {Sukenik}}]{Moses2021}%
  \BibitemOpen
  \bibfield  {author} {\bibinfo {author} {\bibfnamefont {D.}~\bibnamefont
  {Moses}}, \bibinfo {author} {\bibfnamefont {K.}~\bibnamefont {Guadalupe}},
  \bibinfo {author} {\bibfnamefont {F.}~\bibnamefont {Yu}}, \bibinfo {author}
  {\bibfnamefont {E.}~\bibnamefont {Flores}}, \bibinfo {author} {\bibfnamefont
  {A.}~\bibnamefont {Perez}}, \bibinfo {author} {\bibfnamefont {R.~L.}\
  \bibnamefont {McAnelly}}, \bibinfo {author} {\bibfnamefont {N.~M.}\
  \bibnamefont {Shamoon}}, \bibinfo {author} {\bibfnamefont {E.}~\bibnamefont
  {Cuevas-Zepeda}}, \bibinfo {author} {\bibfnamefont {A.}~\bibnamefont {Merg}},
  \bibinfo {author} {\bibfnamefont {E.~W.}\ \bibnamefont {Martin}}, \bibinfo
  {author} {\bibfnamefont {A.~S.}\ \bibnamefont {Holehouse}},\ and\ \bibinfo
  {author} {\bibfnamefont {S.}~\bibnamefont {Sukenik}},\ }\bibfield  {journal}
  {\bibinfo  {journal} {bioRxiv}\ }\href
  {https://doi.org/10.1101/2021.11.24.469609} {10.1101/2021.11.24.469609}
  (\bibinfo {year} {2021})\BibitemShut {NoStop}%
\bibitem [{\citenamefont {Adame-Arana}\ \emph {et~al.}(2020)\citenamefont
  {Adame-Arana}, \citenamefont {Weber}, \citenamefont {Zaburdaev},
  \citenamefont {Prost},\ and\ \citenamefont {J{\"u}licher}}]{AdameArana2020}%
  \BibitemOpen
  \bibfield  {author} {\bibinfo {author} {\bibfnamefont {O.}~\bibnamefont
  {Adame-Arana}}, \bibinfo {author} {\bibfnamefont {C.~A.}\ \bibnamefont
  {Weber}}, \bibinfo {author} {\bibfnamefont {V.}~\bibnamefont {Zaburdaev}},
  \bibinfo {author} {\bibfnamefont {J.}~\bibnamefont {Prost}},\ and\ \bibinfo
  {author} {\bibfnamefont {F.}~\bibnamefont {J{\"u}licher}},\ }\bibfield
  {booktitle} {\emph {\bibinfo {booktitle} {Biophysical Journal}},\ }\href
  {https://doi.org/10.1016/j.bpj.2020.07.044} {\bibfield  {journal} {\bibinfo
  {journal} {Biophys. J.}\ }\textbf {\bibinfo {volume} {119}},\ \bibinfo
  {pages} {1590} (\bibinfo {year} {2020})}\BibitemShut {NoStop}%
\bibitem [{\citenamefont {Hondele}\ \emph {et~al.}(2020)\citenamefont
  {Hondele}, \citenamefont {Heinrich}, \citenamefont {De~Los~Rios},\ and\
  \citenamefont {Weis}}]{Hondele2020}%
  \BibitemOpen
  \bibfield  {author} {\bibinfo {author} {\bibfnamefont {M.}~\bibnamefont
  {Hondele}}, \bibinfo {author} {\bibfnamefont {S.}~\bibnamefont {Heinrich}},
  \bibinfo {author} {\bibfnamefont {P.}~\bibnamefont {De~Los~Rios}},\ and\
  \bibinfo {author} {\bibfnamefont {K.}~\bibnamefont {Weis}},\ }\href
  {https://doi.org/10.1042/ETLS20190190} {\bibfield  {journal} {\bibinfo
  {journal} {Emerg. Top. Life Sci.}\ }\textbf {\bibinfo {volume} {4}},\
  \bibinfo {pages} {343} (\bibinfo {year} {2020})}\BibitemShut {NoStop}%
\bibitem [{\citenamefont {Soeding}\ \emph {et~al.}(2020)\citenamefont
  {Soeding}, \citenamefont {Zwicker}, \citenamefont {Sohrabi-Jahromi},
  \citenamefont {Boehning},\ and\ \citenamefont {Kirschbaum}}]{Soeding2019}%
  \BibitemOpen
  \bibfield  {author} {\bibinfo {author} {\bibfnamefont {J.}~\bibnamefont
  {Soeding}}, \bibinfo {author} {\bibfnamefont {D.}~\bibnamefont {Zwicker}},
  \bibinfo {author} {\bibfnamefont {S.}~\bibnamefont {Sohrabi-Jahromi}},
  \bibinfo {author} {\bibfnamefont {M.}~\bibnamefont {Boehning}},\ and\
  \bibinfo {author} {\bibfnamefont {J.}~\bibnamefont {Kirschbaum}},\ }\href
  {https://doi.org/10.1101/694406} {\bibfield  {journal} {\bibinfo  {journal}
  {Trends Cell Biol.}\ }\textbf {\bibinfo {volume} {30}},\ \bibinfo {pages} {4}
  (\bibinfo {year} {2020})}\BibitemShut {NoStop}%
\bibitem [{\citenamefont {Kirschbaum}\ and\ \citenamefont
  {Zwicker}(2021)}]{Kirschbaum2021}%
  \BibitemOpen
  \bibfield  {author} {\bibinfo {author} {\bibfnamefont {J.}~\bibnamefont
  {Kirschbaum}}\ and\ \bibinfo {author} {\bibfnamefont {D.}~\bibnamefont
  {Zwicker}},\ }\href {https://doi.org/10.1098/rsif.2021.0255} {\bibfield
  {journal} {\bibinfo  {journal} {J. R. Soc. Interface}\ }\textbf {\bibinfo
  {volume} {18}},\ \bibinfo {pages} {20210255} (\bibinfo {year}
  {2021})}\BibitemShut {NoStop}%
\bibitem [{\citenamefont {Klosin}\ \emph {et~al.}(2020)\citenamefont {Klosin},
  \citenamefont {Oltsch}, \citenamefont {Harmon}, \citenamefont {Honigmann},
  \citenamefont {J{\"u}licher}, \citenamefont {Hyman},\ and\ \citenamefont
  {Zechner}}]{Klosin2020}%
  \BibitemOpen
  \bibfield  {author} {\bibinfo {author} {\bibfnamefont {A.}~\bibnamefont
  {Klosin}}, \bibinfo {author} {\bibfnamefont {F.}~\bibnamefont {Oltsch}},
  \bibinfo {author} {\bibfnamefont {T.}~\bibnamefont {Harmon}}, \bibinfo
  {author} {\bibfnamefont {A.}~\bibnamefont {Honigmann}}, \bibinfo {author}
  {\bibfnamefont {F.}~\bibnamefont {J{\"u}licher}}, \bibinfo {author}
  {\bibfnamefont {A.~A.}\ \bibnamefont {Hyman}},\ and\ \bibinfo {author}
  {\bibfnamefont {C.}~\bibnamefont {Zechner}},\ }\href
  {https://doi.org/10.1126/science.aav6691} {\bibfield  {journal} {\bibinfo
  {journal} {Science}\ }\textbf {\bibinfo {volume} {367}},\ \bibinfo {pages}
  {464} (\bibinfo {year} {2020})}\BibitemShut {NoStop}%
\bibitem [{\citenamefont {Deviri}\ and\ \citenamefont
  {Safran}(2021)}]{Devirie2021}%
  \BibitemOpen
  \bibfield  {author} {\bibinfo {author} {\bibfnamefont {D.}~\bibnamefont
  {Deviri}}\ and\ \bibinfo {author} {\bibfnamefont {S.~A.}\ \bibnamefont
  {Safran}},\ }\bibfield  {journal} {\bibinfo  {journal} {Proc. Natl. Acad.
  Sci. USA}\ }\textbf {\bibinfo {volume} {118}},\ \href
  {https://doi.org/10.1073/pnas.2100099118} {10.1073/pnas.2100099118} (\bibinfo
  {year} {2021})\BibitemShut {NoStop}%
\bibitem [{\citenamefont {Bracha}\ \emph {et~al.}(2019)\citenamefont {Bracha},
  \citenamefont {Walls},\ and\ \citenamefont {Brangwynne}}]{Bracha2019}%
  \BibitemOpen
  \bibfield  {author} {\bibinfo {author} {\bibfnamefont {D.}~\bibnamefont
  {Bracha}}, \bibinfo {author} {\bibfnamefont {M.~T.}\ \bibnamefont {Walls}},\
  and\ \bibinfo {author} {\bibfnamefont {C.~P.}\ \bibnamefont {Brangwynne}},\
  }\href {https://doi.org/10.1038/s41587-019-0341-6} {\bibfield  {journal}
  {\bibinfo  {journal} {Nat. Biotechnol.}\ }\textbf {\bibinfo {volume} {37}},\
  \bibinfo {pages} {1435} (\bibinfo {year} {2019})}\BibitemShut {NoStop}%
\bibitem [{\citenamefont {Laan}\ \emph {et~al.}(2015)\citenamefont {Laan},
  \citenamefont {Koschwanez},\ and\ \citenamefont {Murray}}]{Laan2015}%
  \BibitemOpen
  \bibfield  {author} {\bibinfo {author} {\bibfnamefont {L.}~\bibnamefont
  {Laan}}, \bibinfo {author} {\bibfnamefont {J.~H.}\ \bibnamefont
  {Koschwanez}},\ and\ \bibinfo {author} {\bibfnamefont {A.~W.}\ \bibnamefont
  {Murray}},\ }\href {https://doi.org/10.7554/eLife.09638} {\bibfield
  {journal} {\bibinfo  {journal} {eLife}\ }\textbf {\bibinfo {volume} {4}},\
  \bibinfo {pages} {e09638} (\bibinfo {year} {2015})}\BibitemShut {NoStop}%
\bibitem [{\citenamefont {Diepeveen}\ \emph {et~al.}(2018)\citenamefont
  {Diepeveen}, \citenamefont {Gehrmann}, \citenamefont {Pourqui{\'e}},
  \citenamefont {Abeel},\ and\ \citenamefont {Laan}}]{Diepeveen2018}%
  \BibitemOpen
  \bibfield  {author} {\bibinfo {author} {\bibfnamefont {E.~T.}\ \bibnamefont
  {Diepeveen}}, \bibinfo {author} {\bibfnamefont {T.}~\bibnamefont {Gehrmann}},
  \bibinfo {author} {\bibfnamefont {V.}~\bibnamefont {Pourqui{\'e}}}, \bibinfo
  {author} {\bibfnamefont {T.}~\bibnamefont {Abeel}},\ and\ \bibinfo {author}
  {\bibfnamefont {L.}~\bibnamefont {Laan}},\ }\href
  {https://doi.org/10.1093/gbe/evy121} {\bibfield  {journal} {\bibinfo
  {journal} {Genome Biology and Evolution}\ }\textbf {\bibinfo {volume} {10}},\
  \bibinfo {pages} {1765} (\bibinfo {year} {2018})}\BibitemShut {NoStop}%
\bibitem [{\citenamefont {Brauns}\ \emph {et~al.}(2020)\citenamefont {Brauns},
  \citenamefont {I{\\textasciitilde n}igo de~la Cruz}, \citenamefont {Daalman},
  \citenamefont {de~Bruin}, \citenamefont {Halatek}, \citenamefont {Laan},\
  and\ \citenamefont {Frey}}]{Brauns2020a}%
  \BibitemOpen
  \bibfield  {author} {\bibinfo {author} {\bibfnamefont {F.}~\bibnamefont
  {Brauns}}, \bibinfo {author} {\bibfnamefont {L.~M.}\ \bibnamefont
  {I{\\textasciitilde n}igo de~la Cruz}}, \bibinfo {author} {\bibfnamefont
  {W.~K.-G.}\ \bibnamefont {Daalman}}, \bibinfo {author} {\bibfnamefont
  {I.}~\bibnamefont {de~Bruin}}, \bibinfo {author} {\bibfnamefont
  {J.}~\bibnamefont {Halatek}}, \bibinfo {author} {\bibfnamefont
  {L.}~\bibnamefont {Laan}},\ and\ \bibinfo {author} {\bibfnamefont
  {E.}~\bibnamefont {Frey}},\ }\bibfield  {journal} {\bibinfo  {journal}
  {bioRxiv}\ }\href {https://doi.org/10.1101/2020.09.09.290510}
  {10.1101/2020.09.09.290510} (\bibinfo {year} {2020})\BibitemShut {NoStop}%
\bibitem [{\citenamefont {J{\"u}licher}\ \emph {et~al.}(2018)\citenamefont
  {J{\"u}licher}, \citenamefont {Grill},\ and\ \citenamefont
  {Salbreux}}]{Julicher2018}%
  \BibitemOpen
  \bibfield  {author} {\bibinfo {author} {\bibfnamefont {F.}~\bibnamefont
  {J{\"u}licher}}, \bibinfo {author} {\bibfnamefont {S.~W.}\ \bibnamefont
  {Grill}},\ and\ \bibinfo {author} {\bibfnamefont {G.}~\bibnamefont
  {Salbreux}},\ }\href {https://doi.org/10.1088/1361-6633/aab6bb} {\bibfield
  {journal} {\bibinfo  {journal} {Rep. Prog. Phys.}\ }\textbf {\bibinfo
  {volume} {81}},\ \bibinfo {pages} {076601} (\bibinfo {year}
  {2018})}\BibitemShut {NoStop}%
\bibitem [{sou(2022)}]{sourcecode}%
  \BibitemOpen
  \href@noop {} {\bibinfo {title} {Project source code,
  \url{https://github.com/zwicker-group/paper-multicomponent-evolution}}}
  (\bibinfo {year} {2022})\BibitemShut {NoStop}%
\end{thebibliography}%

\begin{figure*}
\centering
\includegraphics[width=\textwidth]{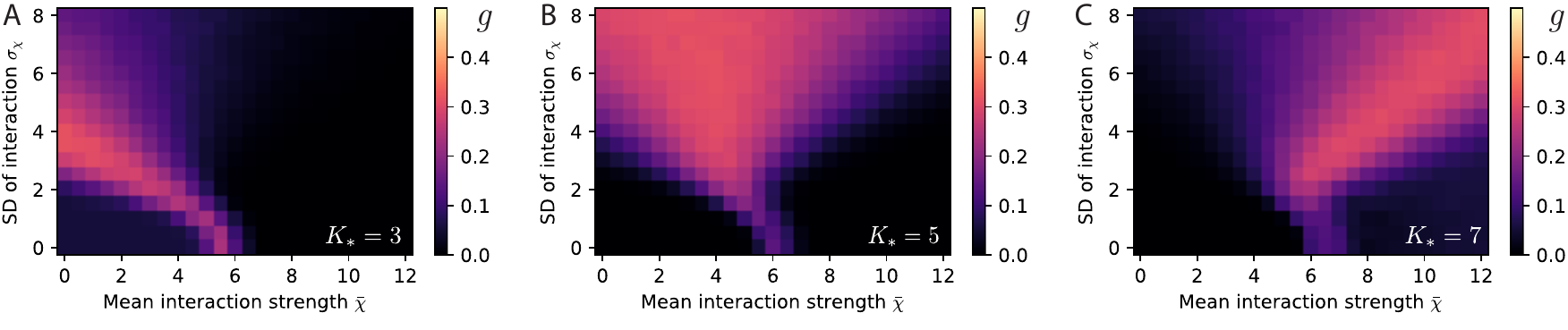}
\caption{Random interaction matrices perform sub-optimally.
(A--C) Performance~$g$ of interaction matrices~$\chi_{ij}$ as a function of their mean $\bar\chi$ and standard deviation $\sigma_\chi$ for $N=9$ components for target phase counts~$K_*=3, 5, 7$ and $w=1$.
}
\label{fig:appendix_random_matrices}
\end{figure*}

\begin{figure*}
\centering
\includegraphics[width=\textwidth]{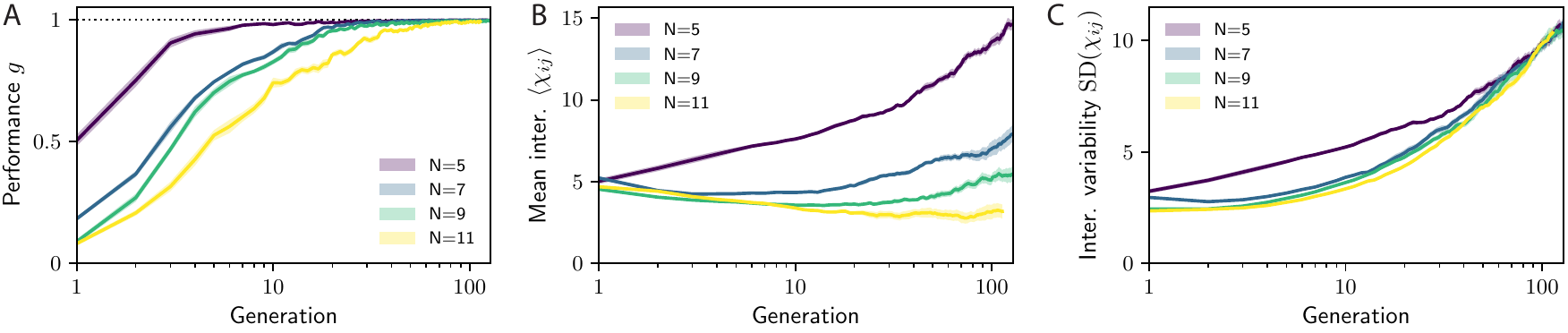}
\caption{Unconstraint evolution obtain optimal solutions at the expense of unphysical large interactions.
(A) Performance $g$ as a function of generation for different number of components~$N$.
(B) Interaction strength~$\mean{\chi_{ij}}$ as a function of generation for various~$N$.
(C) Associated standard deviation~$\mathrm{SD}(\chi_{ij})$ as a function of generation for various~$N$.
(A--C) Additional model parameters are $\sigmaE=1$, $K_*=5$,  and $w=1$.
}
\label{fig:appendix_evolution_unconstraint}
\end{figure*}

\begin{figure}
	\centering	
	\includegraphics[width=\columnwidth]{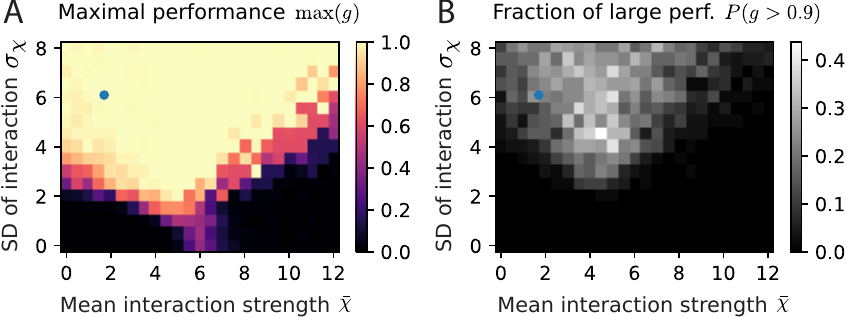}
	\caption{Optimal matrices are frequent in the random ensemble.
	Maximal performance~$g$ (A) and frequency of $g>0.9$ (B) as a function of the mean $\bar\chi$ and standard deviation $\sigma_\chi$ of normally distributed interactions~$\chi_{ij}$.
	The blue dot indicates the statistics of the evolved matrices (\figref{fig:evolution}E).
	$64$ random matrices have been considered for each pair $(\bar\chi, \sigma_\chi)$.
	Additional parameters are $N=9$, $K_*=5$, and $w=1$.
	}
	\label{fig:appendix_monte_carlo}
\end{figure}

\end{document}